\documentclass[aps,pra,twocolumn,groupedaddress,asmsymb,amsmath,amssymb]{revtex4-1}
 \pdfoutput=1
\usepackage{graphicx}
\usepackage{color}
\usepackage{caption}
\usepackage[usenames,dvipsnames]{xcolor}

\newcommand{\be}{\begin{equation}}
\newcommand{\ee}{\end{equation}}
\newcommand{\lb}{\label}

\newcommand{\bj}{{\bf j}}
\newcommand{\bv}{{\bf v}}
\newcommand{\bk}{{\bf k}}
\newcommand{\bx}{{\bf x}}
\newcommand{\br}{{\bf r}}
\newcommand{\bB}{{\bf B}}
\newcommand{\bI}{{\bf I}}
\newcommand{\bJ}{{\bf J}}

\newcommand{\bS}{{\bf S}}

\newcommand{\bomega}{{\mbox{\boldmath $\omega$}}}
\newcommand{\brho}{{\mbox{\boldmath $\rho$}}}

\newcommand{\grad}{{\mbox{\boldmath $\nabla$}}}
\newcommand{\bdot}{{\mbox{\boldmath $\cdot$}}}
\newcommand{\bdots}{{\mbox{\boldmath $:$}}}
\newcommand{\bzed}{{\mbox{\boldmath $0$}}}
\newcommand{\btimes}{{\mbox{\boldmath $\times$}}}

\newcommand{\bbj}{{\bar{\mbox{\boldmath $\j$}}}}
\newcommand{\Dlim}{{{\mathcal D}\mbox{-}\lim}}
\newcommand{\us}{\underline{s}}
\newcommand{\uT}{\underline{T}}
\newcommand{\up}{\underline{p}}
\newcommand{\ubeta}{\underline{\beta}}
\newcommand{\ulambda}{\underline{\lambda}}
\newcommand{\umu}{\underline{\mu}}

\newcommand{\Dto}{{\stackrel{\mathcal{D}}{\longrightarrow}}}

\begin{document}


\title{Cascades and Dissipative Anomalies in Compressible Fluid Turbulence}


\author{Gregory L. Eyink${\,\!}^{1,2}$ and Theodore D. Drivas${\,\!}^1$}
\affiliation{${\,\!}^1$Department of Applied Mathematics \& Statistics, The Johns Hopkins University, Baltimore, MD, USA}
\affiliation{${\,\!}^2$Department of Physics \& Astronomy, The Johns Hopkins University, Baltimore, MD, USA}


\date{\today}

\begin{abstract}
We investigate dissipative anomalies in a turbulent fluid governed by the compressible Navier-Stokes 
equation. We follow an exact approach pioneered by Onsager, which we explain as a non-perturbative application 
of the principle of renormalization-group invariance. In the limit of high Reynolds and P\'eclet numbers, 
the flow realizations are found to be described as distributional or ``coarse-grained'' solutions of 
the compressible Euler equations, with standard conservation laws broken by turbulent anomalies. 
The anomalous dissipation of kinetic energy is shown to be due not only to local cascade, but also 
to a distinct mechanism called pressure-work defect. Irreversible heating in stationary, planar shocks 
with an ideal-gas equation of state exemplifies the second mechanism. Entropy conservation anomalies 
are also found to occur by two mechanisms: an anomalous input of negative entropy (negentropy) 
by pressure-work and a cascade of negentropy to small scales. We derive  ``4/5th-law''-type expressions 
for the anomalies, which allow us to characterize the singularities (structure-function scaling exponents) 
required to sustain the cascades. We compare our approach with alternative theories and empirical 
evidence. It is argued that the ``Big Power-Law in the Sky'' observed in electron density scintillations in 
the interstellar medium is a manifestation of a forward negentropy cascade, or an inverse cascade of usual 
thermodynamic entropy. 
\end{abstract}

\pacs{?????}

\maketitle


\section{Introduction}\label{sec:intro}

Compressible fluids play a vital role in problems of astrophysics 
(interstellar medium \cite{falceta2014turbulence}, star-formation 
\cite{ballesteros2007molecular,federrath2016role}),  applied physics (inertial confinement 
fusion \cite{haines2016laser}), and engineering (high-temperature reactive flows \cite{modest2016chemically}, supersonic aircraft 
design \cite{smits2006turbulent}). Relativistic fluids are necessarily compressible, of course, and occur in astrophysical flows
(pulsars \cite{bucciantini2014review}, gamma-ray bursts \cite{narayan2009turbulent}), 
high-energy physics (heavy-ion collisions \cite{deSouza2016hydrodynamic}), and condensed 
matter physics (graphene \cite{fritz2008quantum,kashuba2008conductivity,muller2009graphene}, 
strange metals \cite{hoyos2013lifshitz,davison2014holographic}). In many of the above examples 
the fluid is either directly observed or indirectly inferred to be in a turbulent state. 
The nature of compressible turbulence has been highly controversial, however.
It is currently debated whether the notion of an ``energy cascade'', as it was developed  
by Kolmogorov \cite{kolmogorov1941local,kolmogorov1941degeneration,kolmogorov1941dissipation}, 
Obukhov \cite{obukhov1941distribution}, 
Onsager \cite{onsager1945distribution,onsager1949statistical}, 
Heisenberg \cite{heisenberg1948theorie} and von Weizs\"acker  \cite{vonweizsacker1948spectre} 
to describe incompressible fluid turbulence, is applicable at all to turbulence in compressible fluids. 
On the one hand,  some authors argue that, much the same as for 
incompressible turbulence, compressible fluids possesses a turbulent inertial range 
``which is immune from direct effects of viscosity and large scale forcing'' \cite{aluie2013scale} through which 
kinetic energy is transferred to small scales by a cascade process that is local in scale. 
On the other hand, exact statistical relations have been derived for non-relativistic compressible 
turbulence \cite{falkovich2010new} and for relativistic turbulence \cite{fouxon2010exact}, which 
do not involve kinetic energy and which have been invoked to argue that 
``the interpretation of the Kolmogorov relation for the incompressible turbulence 
in terms of the energy cascade may be misleading''  \cite{fouxon2010exact}. This is a controversy 
whose resolution has profound consequences for all physical systems where compressible fluid turbulence 
manifests itself.  

The primary physical effect of cascades in incompressible fluids are ``dissipative anomalies," in which 
ideal invariants of the fluid equations, such as kinetic energy, are non-conserved even in the 
inviscid or high Reynolds-number limit. This effect was deduced semi-phenomenologically
from geophysical observations by Taylor \cite{taylor1917observations}
and confirmed in classical wind-tunnel experiments \cite{dryden1943review}. 
For modern evidence from numerical simulations and experiments, see 
\cite{sreenivasan1984scaling,kaneda2003energy,pearson2002measurements}. 
This type of empirical evidence motivated the theories of
Kolmogorov \cite{kolmogorov1941local,kolmogorov1941degeneration,kolmogorov1941dissipation}, 
Obukhov \cite{obukhov1941distribution}, 
Onsager \cite{onsager1945distribution,onsager1949statistical}, 
Heisenberg \cite{heisenberg1948theorie} and von Weizs\"acker  \cite{vonweizsacker1948spectre}.  
A particularly deep contribution was made by Onsager \cite{onsager1949statistical,eyink2006onsager}, 
who argued that turbulent fluids could be described by singular (weak) solutions of 
incompressible Euler equations whose kinetic energy balance equations would be afflicted 
with an anomaly due to the nonlinear cascade mechanism. Onsager's derivation 
was by a (smoothed) version of a point-splitting regularization, which yielded for the 
anomaly an expression closely related to the Kolmogorov 4/5th-law but valid for 
individual flow realizations, without averaging over ensembles \cite{eyink2006onsager}.  
Kolmogorov's weaker statistical relation is, of course, well-known to physicists, e.g. 
Polyakov has pointed out the formal analogy of Kolmogorov's relation and its point-splitting derivation 
to axial anomalies in quantum gauge field theories \cite{polyakov1992conformal,polyakov1993theory}. 
Onsager's deeper contribution has  received little attention in the physics community, on the other 
hand, although the many predictions of Onsager's analysis are consistent with all available experimental evidence.
In particular, his prediction of $1/3$ H\"older singularities for the velocity field has been confirmed 
experimentally (e.g. \cite{kestener2004generalizing}). 
In fact, an entire multifractal spectrum of singularities has been measured,  
as in the later elaboration of Parisi-Frisch \cite{frisch1985singularity,uriel1995turbulence}. 
 
\textcolor{black}{We show here that the Onsager theory carries over to compressible fluids, completing 
earlier work of Aluie \cite{aluie2011compressible,aluie2013scale}.}
There have been extensive further developments of Onsager's ideas, which we shall exploit.
In particular, we follow closely the approaches of Eyink \cite{eyink1994energy,eyink1995local}, 
Constantin et al. \cite{constantin1994onsager}, and Duchon-Robert \cite{duchon2000inertial}, 
who derived necessary conditions for dissipative anomalies of kinetic energy 
in turbulent solutions of incompressible Euler equations. Subsequently, there has been very deep 
mathematical work constructing dissipative, H\"older-continuous Euler solutions for the incompressible case 
by ``convex integration''   methods, using ideas originating in the Nash-Kuiper theorem and Gromov's h-principle 
(e.g. see DeLellis \& Szkeleyhidi \cite{delellis2013continuous,delellis2012h}). 
This circle of ideas led recently to a proof that Onsager's 1/3 
H\"older exponent is sharp \cite{isett2016proof}. These remarks might suggest that a very high level of mathematical sophistication 
is necessary to grasp the essentials of Onsager's  ideas on turbulent weak solutions. This is not the case. As a matter of fact, 
Onsager's ideas are very closely related to standard physical notions of spatial coarse-graining and 
renormalization-group invariance \cite{stueckelberg1951normalization,gellmann1954quantum,bogolyubov1955group}. 
In addition to extending Onsager's approach to compressible fluids 
and deriving many new testable predictions, we shall also explain carefully the connection to renormalization-group 
ideas. By means of this intuitive but rigorous approach, we shall resolve 
the controversies concerning non-relativistic compressible fluid turbulence.  In a companion paper, we 
further extend our analysis to relativistic fluid turbulence \cite{eyink2017cascadesII}. 

Turbulence is an essential strong-coupling problem to which 
perturbation theory does not apply, so that, as in Onsager's original work, some mathematical tools of 
nonlinear analysis must be employed. The required background for full understanding of the finer points is 
mathematical analysis at a theoretical-physics level such as contained in \cite{choquet1982analysis}, 
particularly standard spatial $L^p$-norms (\S I.D.10) and basic theory of distributions/generalized functions
(\S VI.A-B). The tools employed are similar to those in the mathematical theory of 
fluid shock solutions. Most of our analysis can be grasped without even that technical 
background, but assuming just some familiarity with spatial coarse-graining and fluid turbulence.   

\section{Compressible Navier-Stokes and Hydrodynamic Scaling}\label{sec:NS}

The model equations that we employ for (non-relativistic) compressible fluids in this paper are the 
standard Navier-Stokes equations in space dimension $d.$ These govern the evolution of the conserved densities 
(per volume) of mass $\rho,$ momentum $\bj,$ and total energy $E=|\bj|^2/2\rho+u,$ kinetic energy density 
$|\bj|^2/2\rho$ plus internal energy density $u,$ by 
\be \partial_t\rho+\grad\bdot(\rho\bv)=0, \lb{eq1} \ee
\be  \partial_t(\rho\bv)+\grad\bdot(\rho\bv\bv+p\bI-2\eta\bS-\zeta \Theta\bI)=\bzed, \lb{eq2} \ee
\begin{eqnarray} 
&& \partial_t(\frac{1}{2}\rho v^2+u)+\grad\bdot[(u+p+\frac{1}{2}\rho v^2)\bv \cr
&& \hspace{50pt} -\kappa \grad T -2\eta \bS\bdot\bv-\zeta \Theta\bv]=0. 
\lb{eq3} \end{eqnarray} 
A fluid velocity $\bv$ has been defined conventionally 
by $\bv=\bj/\rho,$ which is thus associated to the transport of mass. This is not the only possible 
choice of a fluid velocity. e.g. \cite{onsager1945theories,brenner2005navier}, but it is the most familiar one generally employed
for a non-relativistic fluid. Above, $\eta(u,\rho)$ is the shear viscosity, $\zeta(u,\rho)$ is the bulk viscosity, 
and $\kappa(u,\rho)$ is the thermal conductivity,  and 
\be  S_{ij} = \frac{1}{2}\left(\frac{\partial v_i}{\partial x_j}+\frac{\partial v_i}{\partial x_j}-\frac{1}{d}(\grad\bdot\bv)\delta_{ij}\right),
\quad \Theta = \grad\bdot\bv \lb{eq4} \ee
are the strain tensor and dilatational field, respectively.  
We here admit any thermodynamically consistent relations for the pressure $p(u,\rho)$ (the equation of state) and
for the absolute temperature $T(u,\rho).$ 

 It should be pointed out that this set of equations
has some well-known deficiencies in representing the internal structure of strong shocks, whose thickness
is of the order of the mean-free-path length of the fluid  
\cite{mott1951solution,liepmann1962structure,salomons1992usefulness}. This may cause 
concern, since compressible fluid turbulence is well-known to develop numerous small-scale
``shocklets".  A more fundamental model for the dynamics of a neutral (non-ionized) gas would be the Boltzmann kinetic equation, 
whose solutions agree well with the experimental data for strong shocks. However, we expect that all of our conclusions 
below will still apply if such a kinetic description is employed,  as we shall be concerned with length-scales much greater 
than the width of the shock front where, for both kinetic and fluid models, a similar description emerges as a discontinuous 
weak/distributional solution of the compressible Euler equations \cite{yu2005hydrodynamic}. 
The use of a fluid description from the outset 
greatly simplifies our analysis, but similar arguments should carry over to kinetic theory. We also do not discuss 
here the effects of molecular noise, which for a thermodynamically consistent description of a compressible Navier-Stokes 
fluid requires stochastic PDE's with suitable multiplicative noise given by a fluctuation-dissipation relation 
\cite{morozov1984langevin,eyink1990dissipation}. The effects of such noise are quite significant, presumably leading 
to a ``stochastic anomaly'' in addition to the dissipative anomaly already discussed \cite{eyink1996turbulence,
mailybaev2015stochastic,mailybaev2016spontaneous,mailybaev2016spontaneously}. This is an issue 
of fundamental importance for the problems of predicting, reproducing, or controlling turbulent flows 
\cite{leith1972predictability,ruelle1979microscopic} but, as argued further below, addition of thermal noise 
does not alter our conclusions in this paper on dissipative anomalies   
   
Compressible fluid turbulence is characterized by several dimensionless number groups which are revealed 
by a scaling of the fluid equations. There is more than one way to rescale the equations. Here we follow a very simple 
approach, introducing dimensionless variables
\be  \hat{\rho}=\rho/\rho_0, \quad \hat{\bv}=\bv/v_0, \quad \hat{u}=u/\rho_0 v_0^2 \lb{eq5} \ee 
\be  \hat{\bx}=\bx/L_0, \quad \hat{t}=t/(L_0/v_0) \lb{eq6} \ee
\be  \hat{p} = p/\rho_0 v_0^2, \quad \hat{T}=T/T_0. \lb{eq7} \ee
Above $\rho_0$ and $v_0$ are typical densities and velocities, such as spatial mean or r.m.s. values. 
$L_0$ is the large length-scale of the turbulent flow, e.g. the integral length, or another length
characterizing the scale of injection of kinetic energy either by external stirring or by initial data.
For temperature scale $T_0$ we may also take an r.m.s. value or, alternatively, $T_0=T(\rho_0 v_0^2,\rho_0).$
The non-dimensonalized equations of motion then become
\be  \partial_{\hat{t}}\hat{\rho}+\hat{\grad}\bdot(\hat{\rho}\hat{\bv})=0, \lb{eq8} \ee
\be  \partial_{{\hat t}}(\hat{\rho}\hat{\bv})+\hat{\grad}\bdot(\hat{\rho}\hat{\bv}\hat{\bv}+\hat{p}\bI-2\hat{\eta}\hat{\bS}-\hat{\zeta} \hat{\Theta}\bI)=\bzed, \lb{eq9} \ee
\begin{eqnarray}
&& \partial_{\hat{t}}(\frac{1}{2}\hat{\rho} \hat{v}^2+\hat{u})+\hat{\grad}\bdot[(\hat{u}+\hat{p}+\frac{1}{2}\hat{\rho} \hat{v}^2)\hat{\bv}\cr
&& \hspace{50pt} -\hat{\kappa} \hat{\grad} \hat{T} -2\hat{\eta} \hat{\bS}\bdot\hat{\bv}-\hat{\zeta} \hat{\Theta}\hat{\bv}]=0, 
\lb{eq10} \end{eqnarray} 
where 
\be  \hat{\eta}(\hat{u},\hat{\rho})=\frac{\eta}{\rho_0v_0L_0}, \quad \hat{\zeta}(\hat{u},\hat{\rho})=\frac{\zeta}{\rho_0v_0L_0} \lb{eq11} \ee
are inverse Reynolds numbers associated to the shear and bulk viscosities, respectively, and 
\be  \hat{\kappa}(\hat{u},\hat{\rho})=\frac{\kappa T_0}{\rho_0v_0^3L_0}, \lb{eq12} \ee
is an inverse P\'eclet number. Fully-developed turbulent flow occurs when the Reynolds and P\'eclet numbers
(as functions of $\hat{u}$, $\hat{\rho}$) are uniformly very large compared to unity, and $\hat{\eta},$ $\hat{\zeta},$
$\hat{\kappa}$ are small. It should be borne in mind that these dimensionless transport quantities are generally
made smaller, not by decreasing $\eta,$ $\zeta,$ $\kappa$ but instead typically by increasing $\rho_0,$ $v_0$ or 
especially $L_0.$ We hereafter omit the hats ``$\ \widehat{\cdot} \ $'' on all variables, but always 
assume that non-dimensionalization has been carried out. In particular, when we discuss below the ideal limit 
$\eta,$ $\zeta,$ $\kappa\rightarrow 0$ we always mean more literally that $\hat{\eta},$ $\hat{\zeta},$
$\hat{\kappa}\ll 1.$  
  
There is one other dimensionless variable which plays a very important role in compressible 
turbulence that does not appear explicitly above. This is the Mach number or the inverse of the dimensionless 
sound speed: 
\be  Ma = 1/\hat{c}_s= v_0/c_s,  \lb{eq13} \ee
with $c_s=\sqrt{(\partial p/\partial \rho)_{s_n}}$ the adiabatic sound speed (the density derivative being taken at fixed 
entropy per particle $s_n$). Of course, the Mach number as defined above is also a variable function 
of $\hat{u},$ $\hat{\rho}.$ The properties of compressible turbulence are very strongly dependent upon the 
Mach number. However, the main results of the present work are valid for any Mach number. We shall comment 
below on those parts of our analysis that make any assumptions depending on the Mach number,  either 
explicitly or implicitly. 
 
\section{Dissipative Anomalies}\label{sec:anomaly}

Immediate consequences of the compressible Navier-Stokes equations (\ref{eq1})-(\ref{eq3}) are the kinetic energy balance 
\begin{eqnarray} 
&& \partial_t(\frac{1}{2}\rho v^2)+\grad\bdot\left[(p+\frac{1}{2}\rho v^2)\bv-2\eta \bS\bdot\bv-\zeta \Theta\bv\right]\cr
&&\qquad\qquad =p(\grad\bdot \bv)-2\eta|\bS|^2-\zeta \Theta^2
\lb{eq14} \end{eqnarray}
and the internal energy balance: 
\be  \partial_t u+\grad\bdot\left[u\bv-\kappa\grad T\right]=-p(\grad\bdot \bv)+2\eta|\bS|^2+\zeta \Theta^2, \lb{eq15} \ee
where $2\eta|\bS|^2$, $\zeta \Theta^2$ are the energy dissipation per volume due to shear and bulk viscosity, 
respectively. Because total energy is conserved, the kinetic energy lost is precisely equal to the internal energy gained. 
 
The balance equations analogous to (\ref{eq14}),(\ref{eq15}) above hold also in the limit $Ma\ll 1$ of low Mach numbers. The 
incompressible Navier-Stokes equation then governs the velocity field, in which only the shear viscosity survives. 
The temperature field obeys a passive advection equation with viscous heating as a source term. 
The remarkable empirical fact for incompressible turbulence is that the viscous dissipation per mass 
$Q_\nu=2\nu |\bS|^2,$ where $\nu=\eta/\rho$ is the kinematic viscosity, appears to have a {\it non-vanishing} 
limit as $\nu\rightarrow 0.$ Full documentation of the relevant laboratory experiments and numerical simulations 
can be found in the references \cite{sreenivasan1984scaling,kaneda2003energy,pearson2002measurements}. 
A more precise statement is that a distributional limit 
\footnote{The limit appears to exist in an even stronger sense than distributionally,
which implies limits of $Q_\nu$ exist only when smeared with $C^\infty$, compactly supported test functions.
Experimentally, the limit seems to exist even if $Q_\nu$ is smeared with bounded continuous 
functions (weak convergence of measures) or even when integrated over compact sets, such as balls 
or cubes of fixed size as $\nu\rightarrow 0$. See \cite{meneveau1987multifractal,meneveau1991multifractal}}
of the viscous dissipation field appears to exist 
\be  Q = \Dlim_{\nu\rightarrow 0} Q_\nu \lb{eq16} \ee
and, by experimental evidence, yields in the infinite Reynolds number limit a positive measure $D$ with multifractal 
scaling properties \cite{meneveau1987multifractal,meneveau1991multifractal}. 
As shown in the work of Duchon \& Robert \cite{duchon2000inertial}, this same measure appears as an anomaly term in the kinetic
energy balance equation 
\be  \partial_t(\frac{1}{2} v^2)+\grad\bdot\left[(p+\frac{1}{2} v^2)\bv\right] = -Q \lb{eq17} \ee
for weak solutions of the incompressible Euler equations which are obtained by strong $L^3$ limits of incompressible Navier-Stokes
solutions as $\nu\rightarrow 0.$  Duchon \& Robert \cite{duchon2000inertial} also derived an inertial-range expression 
for $Q$ closely related to the Kolmogorov ``4/5th-law''. This allowed them to prove a refined version of the Onsager 
singularity theorem, namely, that $p$th-order scaling exponents
$\zeta_p^v$ of (absolute) velocity-increments must satisfy $\zeta_p^v\leq p/3$ for $p\geq 3,$ or otherwise $Q\equiv 0.$
See also \cite{constantin1994onsager,eyink1995local}. 
The empirical fact that kinetic energy dissipation has a non-vanishing limit for infinite Reynolds numbers, 
within the accuracy of current measurements, is so central to the modern understanding of incompressible 
fluid turbulence that it is sometimes called the ``zeroth-law of turbulence''. 

The fundamental hypothesis of the present work is that there shall similarly be a non-zero distributional limit 
\be  Q=\Dlim_{\eta,\zeta,\kappa\rightarrow 0} [2\eta|\bS|^2+\zeta \Theta^2] = Q_\eta+ Q_\zeta >0\lb{eq18} \ee
for viscous dissipation in compressible fluid turbulence. A number of previous works have 
investigated the statistical properties of the viscous dissipation in compressible turbulence, 
e.g. \cite{jagannathan2011high,jagannathan2016reynolds,ni2015numerical}. In particular, 
\cite{ni2015numerical} presents direct empirical 
evidence for the ``zeroth-law'' of compressible turbulence. Furthermore, there 
are simple shock solutions of the compressible Navier-Stokes solution which converge 
as $\eta,\zeta,\kappa\rightarrow 0$ strongly (and thus distributionally) to weak solutions 
of compressible Euler equations and for which $Q>0.$  For example, see Appendix A.
In this respect, the theory of compressible fluids is better off than the incompressible theory,
where there are still no rigorous examples of dissipative Euler solutions 
obtained by the physical limit of vanishing viscosity. The purpose of the 
present paper is to develop the consequences of hypothesis (\ref{eq18}) for compressible turbulence.

Furthermore, we shall also consider in this work the balance equation 
for the entropy density per volume which is implied by the compressible Navier-Stokes equation   As is 
well-known (e.g. \cite{degroot2013non}, Ch. XII, \S 1; \cite{landau2013fluid}, \S 49) this 
balance equation has the form   
\begin{eqnarray}
&& \partial_t s +\grad\bdot\left(s\bv -\frac{\kappa\grad T}{T}\right) \cr 
&& \hspace{50pt} = 
\frac{\kappa|\grad T|^2}{T^2} + \frac{2\eta|\bS|^2}{T}+\frac{\zeta \Theta^2}{T}, 
\lb{eq19} \end{eqnarray}
where entropy production on the righthand side is positive, consistent with the second law of 
thermodynamics. \textcolor{black}{Although fluid turbulence is a strongly dissipative macroscopic 
process, there seem to have been remarkably few attempts to understand its consistency with 
the thermodynamic second law. A pioneering work in this direction is a 1949 paper of Obukhov that considered
the entropy balance for incompressible fluid turbulence in the low Mach-number limit 
\cite{obukhov1949temperature}. Extending Obukhov's theory to compressible fluid turbulence 
at arbitrary Mach numbers is one of the principal motivations of our paper.} 

While entropy is conserved for smooth solutions of the compressible Euler 
equations, it is very natural to hypothesize that the entropy balance will also be anomalous for 
compressible turbulent flow and that there will be a non-vanishing limiting measure 
\begin{eqnarray}
\Sigma &=&\Dlim_{\eta,\zeta,\kappa\rightarrow 0}  \left[
\frac{\kappa|\grad T|^2}{T^2} + \frac{2\eta|\bS|^2}{T}+\frac{\zeta \Theta^2}{T}\right] \cr
&=& \Sigma_{therm} + \Sigma_{shear} + \Sigma_{bulk} >0 
\lb{eq20} \end{eqnarray}
describing anomalous entropy production. Simple shock solutions provide examples 
of such entropy anomalies for weak solutions of compressible Euler equations with 
step-discontinuities (Appendix A), but milder H\"older singularities typical of turbulent flow should 
suffice for anomalous entropy production.  In this work we shall derive an inertial-range expression
for $\Sigma$ which shows that an ``inverse cascade'' of entropy can provide a mechanism 
for an entropy anomaly and we characterize the type of H\"older singularities 
of the turbulent solutions required to sustain a non-vanishing entropy flux.


\section{Theoretical Approach}\label{sec:theory}

It is often assumed reflexively that fluid turbulence must be treated probabilistically.
For some problems statistical ensembles are essential, for example, 
for predicting the future of a given turbulent flow 
\cite{leith1972predictability,eyink1996turbulence,mailybaev2016spontaneous,mailybaev2016spontaneously}.
For many problems, however, statistical methods are wholly inadequate, because one always 
observes in Nature a single turbulent flow realization. If one wants to understand the effects of 
turbulence in a specific solar flare event, one does not have the luxury of averaging over an 
ensemble of flares!  An approach that is capable of of treating individual flow realizations is 
intrinsically more fundamental than a probabilistic treatment, because statistical relations can 
always be obtained by subsequent averaging over ensembles. For these reasons, we shall make no 
use of statistical ensembles in the present paper. When we have occasion below to consider 
long-time steady states we shall employ time-averages and global space-averages, denoted 
by $\langle\cdot\rangle,$ which may be operationally obtained in principle from a single flow 
realization. 

Our analysis will be based not on ensemble-averaging but instead upon spatial and/or 
temporal coarse-graining, which we employ as a regularizer. Note that 
the existence of non-vanishing dissipative anomalies as in (\ref{eq18}) and (\ref{eq20})  
requires that gradients of fluid variables must diverge, $|\grad\bv|,|\grad T|\rightarrow\infty$
as $\eta,\zeta,\kappa\rightarrow 0$. This is an ultraviolet divergence due the development 
of high-wavenumber excitations in the ideal limit,  or, as described by Onsager \cite{onsager1945distribution}, 
a ``violet catastrophe''.  It is a consequence of these divergences that the fluid equations (\ref{eq1})-(\ref{eq3}) can 
no longer remain meaningful in the naive sense, because they involve the above diverging gradients. 
To make sense of the dynamics in the ideal limit, Onsager used a point-splitting regularization 
partially smoothed by a filtering kernel \cite{eyink2006onsager}.  We employ an alternative approach 
\cite{constantin1994onsager,eyink1995local} more closely related to renormalization group (RG) methods, 
with fields $f(\bx,t)$ spatially coarse-grained as 
\be  \bar{f}_\ell(\bx,t)=\int d^dr\ G_\ell(\br) f(\bx+\br,t) \lb{eq21} \ee
where $G_\ell(\br)=\ell^{-d}G(\br/\ell)$ and the filter kernel $G$ is non-negative, smooth, rapidly 
decreasing in space, and normalized so that $\int d^d r \ G(\br) =1.$ 
This coarse-graining operation is a natural regularization which removes short-distance 
divergences. This can be seen from Cauchy-Schwartz bounds on the coarse-grained gradients:  
\be 
|\grad\bar{f}_\ell|\leq (1/\ell)\sqrt{\int d^dr\ |(\grad G)_\ell(\br)|^2\cdot\int_{{\rm supp}(G_\ell)} d^dr\ f^2(\bx+\br)},
\lb{eq22} \ee
which are finite as long as the function $f$ is locally square-integrable. 
Such estimates are intuitively obvious, because high-wavenumbers $k\gtrsim 1/\ell$ have 
been removed. As we shall see, the coarse-graining regularization in (\ref{eq21})  is more powerful and more general 
than the point-splitting  originally used by Kolmogorov and Onsager \footnote{Here we may note that the 
coarse-graining regularization permits one to analyze dissipative anomalies as well in Lagrangian conservation
laws such as fluid circulations  \cite{eyink2006turbulent} or magnetic fluxes \cite{eyink2006turbulent} 
for advected loops, but it is very unclear how to apply point-splitting to Lagrangian invariants.}. 
The coarse-grained field in (\ref{eq21}) is analogous to a ``block-spin'' in a Wilson-Kadanoff RG scheme 
\cite{kadanoff1966scaling,wilson1971renormalization}. A crucial point is that the coarse-graining 
is a purely passive operation, which corresponds simply to ``taking off one's spectacles". Although it smooths 
out divergences, no objective physical phenomenon can depend upon the arbitrary scale $\ell$ of coarse-graining.
We shall below draw important conclusions from this independence, which is a type of non-perturbative 
RG invariance. To keep our notations simple, we shall denote $\bar{f}_\ell$ just as $\bar{f},$
unless it is essential that the length-scale $\ell$ be made explicit. 

Not only do the equations of motion of coarse-grained fluid variables remain well-defined  in the ideal limit, 
but another crucial fact is that dissipative transport in those equations due to the molecular coefficients 
becomes negligible at fixed length scale $\ell$ in the limit $\eta,\ \zeta,\ \kappa\rightarrow 0$. Because 
this is an essential point, we here give a simple proof of this fact. 
Starting with momentum transport, a coarse-graining of the momentum balance equation yields
\be  \partial_t\left(\overline{\rho\bv}\right)+
\grad\bdot\left(\overline{\rho\bv\bv+p\bI-2\eta\bS-\zeta \Theta\bI}\right)=\bzed, \lb{eq23} \ee
because coarse-graining commutes with space and time derivatives. 
The contribution of the shear viscosity can be bounded pointwise using the Cauchy-Schwartz inequality 
\begin{eqnarray} 
&& \left|\grad\bdot \overline{2\eta\bS}(\bx,t)\right|= \frac{2}{\ell}\left|\int d^3r\, (\grad G)_\ell(\br) \bdot \eta(\bx+\br,t) \bS(\bx+\br,t)\right| \cr 
&&\leq  \frac{2}{\ell} \sqrt{ \langle \eta(\bx)\rangle_\ell \times \int d^3r\, |(\grad G)_\ell(\br)|^2 \ \eta(\bx+\br,t) |\bS(\bx+\br,t)|^2} \cr
&& \lb{eq24} \end{eqnarray} 
with $ \langle \eta(\bx,t)\rangle_\ell=\int_{{\rm supp}\ G_\ell} d^3r\, \eta(\bx+\br,t)$ and with ${\rm supp}(G_\ell)$
the compact support set of the function $G_\ell.$ The inverse power $1/\ell,$
arose after using integration by parts to move the gradient to the filter function $G.$ It shows that 
this term is essentially ``irrelevant'' in the RG sense and is damped out for increasing $\ell.$ When $G$ 
is a test function in the Schwartz distribution theory ($G\in C^\infty$ and compactly supported), then so is 
$|\grad G(\cdot-\bx)|^2$ and thus 
\begin{eqnarray}
&& \lim_{\eta,\zeta,\kappa\rightarrow 0} \int d^3r\, |(\grad G)_\ell(\br)|^2 \eta(\bx+\br,t) |\bS(\bx+\br,t)|^2\cr
&& \qquad =  \int  |(\grad G)_\ell(\br-\bx)|^2 \, Q_\eta(d\br) 
\lb{eq25} \end{eqnarray}
by our fundamental hypothesis. On the other hand, $ \langle \eta(\bx)\rangle_\ell\rightarrow 0$ whenever 
$\eta$ tends to zero locally in $L^1.$ An identical argument shows also that 
$ \left|\grad\overline{\zeta\Theta}(\bx,t)\right|\rightarrow 0$  pointwise for fixed length scale $\ell$ 
when $\zeta$ tends to zero locally in $L^1.$ It follows that all of the molecular transport terms in 
the coarse-grained momentum balance become negligible in the limit of high Reynolds numbers.  

This leads to the concept of the ``inertial range'', or the length-scales $\ell$ sufficiently large 
that the molecular transport can be ignored relative to the large-scale momentum transport 
$\sim \rho_0 v_0^2/L_0.$ The previous upper bound shows that this range extends down to at least 
$\ell \sim L_0/\sqrt{Re_{s}}$ with a Reynolds number defined by $1/Re_{s}=\eta_0 Q_\eta/\rho_0^2 v_0^4,$ which is analogous to the 
``Taylor microscale'' of incompressible fluid turbulence. Here we have assumed that $\zeta\sim \eta,$ otherwise 
one must consider also the limit set by $\ell \gtrsim L_0/\sqrt{Re_{b}}$ with another 
``bulk-viscosity Reynolds number'' defined by $1/Re_{b}=\zeta_0 Q_\eta/\rho_0^2 v_0^4.$
It should be emphasized that the above estimates of length-scales where viscosity effects 
become significant are  expected to be over-estimates, because they are deduced from 
mathematical upper bounds on the molecular transport. The range of scales $\ell$ where viscosity  
is significant is usually termed the ``dissipation range'' and extends down to scales of order 
the mean-free path length $\lambda_{mfp}$ of the fluid, where the hydrodynamic description breaks down.

The same arguments apply also to the 
energy balance, where the shear-viscosity contribution is bounded by 
\begin{eqnarray}
&& \left|\grad\bdot \overline{2\eta\bS\bdot\bv}(\bx,t)\right|
\leq   \frac{2}{\ell} \sqrt{\int_{{\rm supp\ G_\ell}} d^3r\, |\bv(\bx+\br)|^2 \eta(\bx+\br,t)} \cr 
&& \qquad \times \sqrt{\int d^3r\, |(\grad G)_\ell(\br)|^2 \ \eta(\bx+\br,t) |\bS(\bx+\br,t)|^2}
\lb{eq26} \end{eqnarray} 
and this vanishes at fixed $\ell,$ for example, if $\bv$ is locally $L^2$ and if $\eta$ tends to zero 
locally in $L^\infty.$ The energy transport by shear viscosity is negligible compared with large-scale advective 
transport $\sim \rho_0 v_0^3/L_0$ again down to length scale $\ell \sim L_0/\sqrt{Re_{s}}$ (at least). 
Similar arguments apply to the other molecular contributions to energy transport. That from bulk viscosity 
tends to zero if $\zeta\rightarrow 0$  locally in $L^\infty$ and is negligible down to at least the length-scale 
$\ell \sim L_0/\sqrt{Re_{b}}.$ Finally, the contribution from thermal conductivity vanishes if temperature $T$ 
is locally $L^2$ and $\kappa\rightarrow 0$ locally in $L^\infty,$ and it may be neglected down at least to 
length-scale $\ell \sim L_0/\sqrt{Pe_{c}}$ for the thermal P\'eclet number defined 
by $1/Pe_{c}=\kappa_0 T_0^2 \Sigma_{therm} /\rho_0^2 v_0^6.$ 

The final conclusion of this argument is that for sufficiently large length-scales $\ell$ (or for all 
$\ell$ in the ideal limit $\eta,\ \zeta,\ \kappa\rightarrow 0$) the following set of coarse-grained balance 
equations hold:
\be \partial_t\overline{\rho}+\grad\bdot(\overline{\rho\bv})=0, \lb{eq27} \ee
\be  \partial_t(\overline{\rho\bv})+\grad\bdot(\overline{\rho\bv\bv+p\bI})=\bzed,\lb{eq28} \ee
\be  \partial_t\Big(\overline{\frac{1}{2}\rho v^2+u}\Big)+\grad\bdot\Big[\overline{(u+p+\frac{1}{2}\rho v^2)\bv}\Big]=0, \lb{eq29} \ee
in which the molecular transport terms are absent. This set of equations for all $\bx,t$ and $\ell$ is mathematically 
equivalent to the statement that any limiting fields $\rho,\bv, u$ are distributional or ``weak" solutions
of the compressible Euler equations (see \cite{eyink2015turbulent}, section 4; \cite{drivas2017onsager}). 
It must be appreciated that this notion of ``distributional/weak solution'' is quite distinct from 
the statement that either the fields $\rho,\bv, u$ or their coarse-grained versions satisfy compressible Euler 
equations in the usual naive sense \footnote{The failure to appreciate this point has been the source of 
many misunderstandings. For example, consider the following typical quote:
\begin{quotation}
\noindent ``We therefore conclude that, for the large eddies which are the 
basis of any turbulent flow, the viscosity is unimportant and may be equated to zero, so that the motion of 
these eddies obeys Euler's equation. In particular, it follows from this that there is no appreciable dissipation 
of energy in the large eddies.'' -- Landau \& Lifschitz \cite{landau1959fluid}, \S 31. 
\end{quotation}
This statement is correct, if one understands it to mean that the {\it viscous} dissipation is negligible
for the large eddies. However, the coarse-graining which permits one to neglect viscosity at large-scales
generates new stresses which do not conserve the energy of the large eddies!}. 
This point can be made clearly by introducing the density-weighted 
Favre-average\footnote{The use of Favre-average cumulants rather than cumulants for the 
original spatial coarse-graining is not essential. Their use does, however, reduce the number 
of additional cumulant terms that appear and permits a simple physical interpretation of each 
such term. For these reasons, Favr\'e-averaging has been very popular in the practical 
engineering modelling of compressible turbulence \cite{garnier2009large}.} \cite{favre1969statistical}. 
\be \widetilde{f}=\overline{\rho f}/\overline{\rho} \lb{eq30} \ee 
and using the expansion of average products $\widetilde{f_1 \cdots f_n}$ into a finite sum of 
$p$th-order cumulants $\tilde{\tau}(f_{i_1},...,f_{i_p}):$
\be \widetilde{f_1 \cdots f_n} =\sum_{I} \prod_{r=1}^{r_I} \tilde{\tau}(f_{i_1^{(r)}},...,f_{i_{p_r}^{(r)}})\lb{eq31} \ee
where the sum is over all distinct partitions $I$ of $\{1,...,n\}$ into $r_I$ disjoint subsets 
$\{i_1^{(r)},...,i_{p_r}^{(r)}\},$ $r=1,...,r_I,$ so that $\sum_{r=1}^{r_I} p_r=n$ for each partition $I$
\cite{huang1987statistical,germano1992turbulence}. One may likewise expand averaged products $\overline{f_1\cdots f_n}$ into 
cumulants $\overline{\tau}(f_1,...,f_n)$ for the original (non-density weighted) spatial coarse-graining. 
Exploiting these cumulant expansions in the ideal balance equations yields an equivalent 
set of equations 
\be \partial_t\bar{\rho}+\grad\bdot(\bar{\rho}\tilde{\bv})=0, \lb{eq32} \ee
\be  \partial_t(\bar{\rho}\tilde{\bv})+\grad\bdot(\bar{\rho}\tilde{\bv}\tilde{\bv}+\bar{\rho}\tilde{\tau}(\bv,\bv)+\bar{p}\bI)=\bzed,\lb{eq33} \ee
\begin{eqnarray}
&& \partial_t\Big(\frac{1}{2}\bar{\rho} |\tilde{\bv}|^2+\frac{1}{2}\bar{\rho}\tilde{\tau}(v_i,v_i)+\bar{u}\Big) 
+\grad\bdot\Big[(\bar{u}+\bar{p})\bar{\bv} +\bar{\tau}(u+p,\bv) \cr
&& \qquad  +\frac{1}{2}\bar{\rho} |\tilde{\bv}|^2\tilde{\bv}+\bar{\rho} \tilde{v}_i\tilde{\tau}(v_i,\bv) 
+\frac{1}{2}\bar{\rho} \tilde{\tau}(v_i,v_i,\bv) \Big]=0.
\lb{eq34} \end{eqnarray}
It is immediately clear that the coarse-grained fields $\bar{\rho},\tilde{\bv}=\bbj/\bar{\rho},$ and $\bar{u},$
although smooth and with all derivatives well-defined, do not satisfy the compressible Euler equations 
in the standard sense and that there are new transport terms at length-scale $\ell$ which were introduced 
by the coarse-graining. It is, of course, not surprising that the effective equations for ``block-spin'' variables 
are renormalized and contain new, complex terms. Note, in particular, that the coarse-graining 
cumulants of second and higher orders are not  simple closed functions of the ``resolved'' fields 
$\bar{\rho},\tilde{\bv},$ $\bar{u}.$ The cumulants are instead very complex functions of the resolved
fields, with non-polynomial non-linearity and non-Markovian dependence on the past history. 
In fact, these cumulants cannot in principle be fixed, deterministic 
functions of the resolved fields, but must be considered stochastic \footnote{One source of such stochasticity 
is thermal noise due to molecular degrees of freedom, so far neglected in our analysis, but other unknown perturbations
can also lead to randomness. This stochasticity 
does not contradict our earlier claim of a ``deterministic approach''. A given weak Euler solution 
$\rho,$ $\bv,$ $u$ over a particular time-interval corresponds to a particular realization of these cumulants. 
It is only if one wants to predict or control the future behavior outside that particular time interval that one must 
recognize the intrinsic stochasticity. Here we may note that vanishingly small stochastic perturbations do not 
alter our conclusions in Section \ref{sec:theory} regarding the description of coarse-grained variables in the inertial-range
by ``weak Euler solutions''. This argument may be made precise for thermal noise by appealing to the 
Onsager fluctuation principle 
\cite{onsager1953fluctuations,graham1978path,eyink1990dissipation}. The latter principle 
states that the probability of observing a particular set of fields $\rho,$ $\bv,$ $u$ as thermal fluctuations is 
related to the additional dissipation/entropy production required to produce the fluctuation. 
See \cite{eyink1990dissipation}, section 4, example 2, 
for compressible Navier-Stokes fluids. 
Even when this excess dissipation is non-vanishing, the noise terms vanish in the coarse-grained equations for the limit 
$\eta,$ $\zeta$ $\kappa\rightarrow 0.$ Details will be given elsewhere, but the argument is essentially the same 
as for the deterministic fluid equations in Section \ref{sec:theory}.  Thus, coarse-grained variables in the inertial-range range are described 
by weak Euler solutions even in the presence of thermal noise.} variables because of their dependence 
on the unknown degrees of freedom below length-scale $\ell$ \cite{eyink1996turbulence}.   In the ``large-eddy simulation'' 
(LES) methodology of turbulence modelling, one seeks computationally tractable closed models of these
cumulant terms as functions of the resolved fields (see \cite{meneveau2000scale,garnier2009large,schmidt2015large}).
As we shall see, the cumulant terms that appear in these 
coarse-grained equations are the source of turbulent cascade and dissipative anomalies for 
weak Euler solutions \cite{eyink1995local}. 

The above description of weak solutions is somewhat novel and designed to make clear 
the close connection with renormalization-group methodology. A more traditional account 
follows by first taking the ideal limit $\eta,\ \zeta,\ \kappa\ \rightarrow 0$ of the coarse-grained 
conservation equations, just as above, and then followed by the limit $\ell\rightarrow 0$.
The coarse-grained balance equations in this order of limits converge in the sense of distributions to 
\be \partial_t\rho+\grad\bdot(\rho\bv)=0, \lb{eq35} \ee
\be  \partial_t(\rho\bv)+\grad\bdot(\rho\bv\bv+p\bI)=\bzed,\lb{eq36} \ee
\be  \partial_t(\frac{1}{2}\rho v^2+u)+\grad\bdot[(u+p+\frac{1}{2}\rho v^2)\bv]=0. \lb{eq37} \ee
This system follows because all space and time derivatives can be transferred to the test functions 
and all coarse-grained fields inside the derivatives converge to their fine-grained 
values under modest assumptions on the fields (e.g. if they are bounded, measurable functions). 
Equivalently, all of the coarse-graining cumulants of the fields converge to zero. 
In contrast to the regularized systems of equations (\ref{eq27})-(\ref{eq29}) or (\ref{eq32})-(\ref{eq34}) where all derivatives are taken 
in the classical sense, in the above set of $\ell\rightarrow 0$ limit equations (\ref{eq35})-(\ref{eq37}) the 
derivatives must be interpreted distributionally, since the limit functions $\rho,$ $\bv,$ $u$
are not generally even once-differentiable. (See further discussion on fluid singularities below.) 
This more conventional notion of weak solution is a suitable mathematical idealization of infinite 
Reynolds-number turbulence, where the inertial range extends to infinitesimally small scales.    
The concept goes back to Onsager \cite{onsager1949statistical}, who termed it ``ideal turbulence.''   

As we now discuss, standard consequences of the Euler equations for smooth ``strong"  
solutions do not generally hold for weak solutions, which are instead afflicted with 
dissipative anomalies due to turbulent cascade. First, we make an important
comment on notations. Whenever coarse-grained quantities marked with $\overline{(\cdots)}$
or $\widetilde{(\cdots)}$ appear hereafter, we shall assume that the ideal limit  
$\eta,\ \zeta,\ \kappa\ \rightarrow 0$ has been taken, unless indicated otherwise
(e.g. by explicitly taking this limit, or by retaining terms with explicit dependence on $\eta,\ \zeta,\ \kappa$). 
This convention for coarse-grained quantities simplifies the expressions involved 
by eliminating the terms which vanish in the ideal limit by the arguments given above.

\section{Energy Cascade}\label{nonrel-energy}\label{sec:energy} 

\subsection{Kinetic Energy}\label{sec:kinetic}

We begin with kinetic energy cascade.  Because $|\bv|^2$ is a convex function of $\bv,$ one has 
\be \frac{1}{2}\bar{\rho}|\tilde{\bv}|^2\leq  \frac{1}{2}\bar{\rho}\widetilde{|\bv|^2} = \frac{1}{2}\overline{\rho|\bv|^2}.\lb{eq38} \ee
Thus, the integral over space of $\frac{1}{2}\bar{\rho}|\tilde{\bv}|^2$ is less than the total kinetic energy, and represents only the 
``resolved'' kinetic energy, while the 2nd-order Favr\'e cumulant 
\be \frac{1}{2}\bar{\rho}\tilde{\tau}(v_i,v_i)= \frac{1}{2}\bar{\rho}[\widetilde{v_i^2} - \tilde{v}_i^2] \geq 0 \lb{eq39} \ee 
represents the ``unresolved'' or ``subscale'' kinetic energy.  In a decaying turbulence without external forcing, the fine-grained 
kinetic energy integrated over space decreases because of the effect of viscosity. Since 
\be  \frac{1}{2} \int d^dx\ \bar{\rho}|\tilde{\bv}|^2 \leq \frac{1}{2}\int d^dx\ \rho|\bv|^2, \lb{eq40} \ee
this decrease must also occur with increasing time for the resolved kinetic energy, despite the negligible
effect of viscosities for $\ell$ in the inertial range. Physically speaking, the kinetic energy will decay whether 
an observer is ``wearing spectacles'' or not. The question thus arises: how can the resolved 
kinetic energy decay, if not through the influence of viscosity? A similar question arises 
for forced-steady states. If the fluid is stirred by a large-scale acceleration field, then it 
is not hard to show that the input of resolved kinetic energy is nearly independent of $\ell$ 
for $\ell\lesssim L,$ the length-scale of the acceleration field (\cite{aluie2013scale}, Appendix B). 
What mechanism is available at length-scales $\ell$ in the inertial-range in order to dispose
of the fixed mechanical power input and to maintain a mean steady-state energy? 
  
\textcolor{black}{An obvious answer} is that the cumulant term in the equation (\ref{eq33}), or the ``subscale stress''
$\tilde{\tau}(\bv,\bv)$, provides an effective dissipation of kinetic energy for $\ell$ in the inertial range. 
This can be seen from the balance equation for the resolved kinetic energy, which may be easily 
calculated from (\ref{eq33}) to be \textcolor{black}{\cite{aluie2011compressible,aluie2013scale}:}
\begin{eqnarray}
&& \partial_t(\frac{1}{2}\bar{\rho} |\tilde{\bv}|^2)+\grad\bdot\left[(\bar{p}+\frac{1}{2}\bar{\rho} |\tilde{\bv}|^2)\tilde{\bv}
+\bar{\rho}\tilde{\tau}(\bv,\bv)\cdot\tilde{\bv}- \frac{\bar{p}}{\bar{\rho}}\bar{\tau}(\rho,\bv) \right] \cr
&& \qquad =\bar{p}(\grad\bdot \bar{\bv})-\frac{\grad\bar{p}}{\bar{\rho}}\bdot\bar{\tau}(\rho,\bv)
+\rho\grad\tilde{\bv}\ \bdots\ \tilde{\tau}(\bv,\bv).
\lb{eq41} \end{eqnarray} 
The final term on the righthand side is the so-called ``deformation work'', or the work done by 
the large-scale velocity-gradient $\grad\tilde{\bv}$ acting against the subscale stress $\tilde{\tau}(\bv,\bv)$.
It thus a typical ``flux-like term'' describing an interaction between resolved and unresolved degrees of freedom 
and, on average, transferring kinetic energy from resolved to unresolved modes. This term is one 
of the main contributors to kinetic energy cascade. It is more traditional to combine the first two terms on 
the righthand side of (\ref{eq41}) into a single term $\bar{p}(\grad\bdot \tilde{\bv})$ \cite{garnier2009large}, 
together with dropping the last term  in the square bracket (representing space-transport of kinetic energy) 
on the lefthand side.  Above we have followed Aluie 
\textcolor{black}{\cite{aluie2011compressible,aluie2013scale}} in separating the contributions 
of resolved pressure-work $\bar{p}(\grad\bdot \bar{\bv})$ and ``baropycnal work" 
$-\grad\bar{p}\bdot\bar{\tau}(\rho,\bv)/\bar{\rho},$  using the simple relation 
\be  \tilde{\bv} = \bar{\bv} + \frac{1}{\bar{\rho}} \bar{\tau}(\rho,\bv). \lb{eq42} \ee
This division was motivated physically in \textcolor{black}{\cite{aluie2011compressible,aluie2013scale}}, 
which pointed out that the resolved pressure work 
is a purely large-scale quantity, whereas the baropycnal work is ``flux-like'' and describes an interaction between the 
resolved pressure-gradient and subscale mass transport. For more discussion of the physics of this term, 
see \textcolor{black}{\cite{aluie2011compressible,aluie2013scale}}. 
The baropycnal work is thus an additional contributor to kinetic energy cascade, 
with total inertial-range energy flux represented by the combination 
\be  Q^{flux}_\ell= \frac{\grad\bar{p}}{\bar{\rho}}\bdot\bar{\tau}(\rho,\bv)
-\bar{\rho}\grad\tilde{\bv}\ \bdots\ \tilde{\tau}(\bv,\bv).  \lb{eq43} \ee
As we shall see presently,
there are also compelling mathematical reasons to make the above separation of the pressure-work. 

The cascade terms in the equation (\ref{eq41}) are \textcolor{black}{a possible} source of the dissipative anomaly 
of kinetic energy for the weak solutions of Euler obtained in the limit first $\eta,\ \zeta, \ \kappa\rightarrow 0$ and 
then $\ell\rightarrow 0.$ Taking the limit $\ell\rightarrow 0$ of the balance equation (\ref{eq41}), one obtains 
\be  \partial_t(\frac{1}{2}\rho v^2)+\grad\bdot[(p+\frac{1}{2}\rho v^2)\bv]=p\circ \Theta-Q_{flux} \lb{eq44} \ee
which is the kinetic energy balance for the limiting weak Euler solution. Here we defined
\be  p\circ \Theta = {\mathcal D}\mbox{-}\lim_{\ell\rightarrow 0}\bar{p} \cdot \bar{\Theta} \lb{eq45} \ee
with $\Theta=\grad\bdot\bv$ and 
\be 
Q_{flux}={\mathcal D}\mbox{-}\lim_{\ell\rightarrow 0} Q^{flux}_\ell,  \lb{eq46} \ee
where ${\mathcal D}\mbox{-}\lim$ denotes limit in the sense of distributions. We now discuss the physical 
meaning of these two terms. 

Because $p$ and $\bv$ are not generally smooth functions, the divergence $\Theta$ exists only as a distribution 
and its product with the non-smooth function $p$ is thus ill-defined and ambiguous. The limit $p\circ \Theta$ 
in (\ref{eq45}) above is a standard approach to define a generalized product of distributions 
\cite{Oberguggenberger92}
and our circle notation ``$\circ$'' is meant to emphasize that this product must be carefully defined. 
Despite this subtlety, however, such a term is exactly the same as that which appears for a smooth 
Euler solution. It clearly represents pressure-work in the large-scales which converts energy 
from mechanical to internal, and vice-versa. Tendency to equipartition of total energy suggests that,  
when the turbulence is mechanically forced, the transfer will be on average from mechanical to internal. 
It was already argued in \textcolor{black}{\cite{aluie2011compressible,aluie2013scale}} 
that the  mean transfer $\langle \bar{p}_\ell \bar{\Theta}_\ell\rangle$ 
at length-scales $>\ell$ will saturate to a constant negative value 
as $\ell$ decreases through the inertial range, and this saturation has been verified in numerical
simulations of subsonic and transonic compressible turbulence 
\cite{aluie2012conservative,ni2015effects}. 
Our mathematical analysis implies that $\langle p\circ \Theta \rangle<0$ will give the saturated level. 

The additional term $Q_{flux}$ that appears in (\ref{eq44}) is, on the other hand, entirely 
missing for smooth Euler solutions and represents a kinetic energy anomaly. It 
is due to the loss of kinetic energy by turbulent cascade to infinitesimally small scales. 
As shown by Onsager \cite{onsager1949statistical,eyink2006onsager} for the case of 
incompressible fluid turbulence, the non-vanishing of such a cascade term requires 
singularities of the fluid fields $\rho,\bv,$ and $p.$ For a complete proof of the analogous result 
for compressible fluids, see  
\textcolor{black}{ the works of Aluie \cite{aluie2011compressible,aluie2013scale,aluieunpub} and} 
the companion paper \cite{drivas2017onsager}. Briefly, the result follows 
by expanding Favre averages into cumulants $\overline{\tau}(f_1,...,f_n)$ of the original (non-density
weighted) coarse-graining. A fundamental fact is that such cumulants and their 
spatial-gradients can be written entirely in terms of space-increments 
$\delta f_i(\br;\bx)= f_i(\bx+\br)- f_i(\bx)$ of the fields $f_i$ (for a proof see 
\cite{eyink2015turbulent}, Appendix B, or \cite{eyink2010notes}). From these basic identities one can derive estimates of the form  
\be  \grad\tilde{\bv} = \frac{O(\delta v)}{\ell} 
\left[ 1+ O\left(\frac{\delta\rho}{\rho}\right) + O\left(\left(\frac{\delta\rho}{\rho}\right)^2\right)\right], \lb{eq47} \ee
\be  \tilde{\tau}(\bv,\bv) = O(\delta v)^2 
\left[ 1+ O\left(\frac{\delta\rho}{\rho}\right) + O\left(\left(\frac{\delta\rho}{\rho}\right)^2\right)\right],  \lb{eq48} \ee
where $\delta v,$ $\delta \rho$ denote increments over the length-scale $\ell.$ 
\textcolor{black}{For example, see eqs.(1)-(2) and intervening relations in the paper of Aluie \cite{aluie2011compressible}.} 
Substituting such expressions 
into the formula for the deformation work yields an analogue for compressible turbulence of the 
``4/5th-law'' of Kolmogorov \cite{kolmogorov1941dissipation}, which expresses the energy flux 
in terms of increments of the basic fields \footnote{\textcolor{black}{In fact, unpublished work of Aluie 
\cite{aluieunpub}} shows how to recover the traditional 4/5th-law of Kolmogorov 
from such coarse-grained expressions for kinetic energy flux.}.  One can easily see that when 
the fields $\bv$ and $\rho$ are space-differentiable, then  $\delta v, \delta\rho=O(\ell),$ and thus 
energy flux due to deformation work vanishes at least  as $O(\ell^2)$ for $\ell\rightarrow 0.$ In order
to sustain a non-vanishing energy flux, the fluid variables must appear ``rough'' for $\ell$ in the 
inertial range. A more precise statement is that the scaling exponents $\zeta_p^v$ of the 
$q$th-order (absolute) velocity structure functions \cite{uriel1995turbulence} must be sufficiently small. 
For example, when density $\rho$ is bounded away from zero and infinity, then non-vanishing of 
$\lim_{\ell\rightarrow 0} \bar{\rho}\grad\tilde{\bv}\ \bdots\ \tilde{\tau}(\bv,\bv)$ requires
\be  \zeta_q^v \leq q/3, \quad \forall q\geq 3, \lb{eq49} \ee
where $q/3$ is the dimensional Kolmogorov value. See \cite{aluieunpub,drivas2017onsager}. 
This is an exact singularity statement for the velocity field 
in compressible turbulence, consistent with possible spatial intermittency. It is interesting 
to note that for structure-function exponents in the range $0<\zeta_q^v,\zeta_q^\rho <p,$ 
the deformation work can be shown to be scale-local \textcolor{black}{{\cite{eyink2005locality,aluie2011compressible}}}
and thus ``cascade'' is an apt description of the transfer process 
\footnote{A caveat has to do with the contribution of the density.
If, as we have assumed in this work, the density is a bounded function and Hoelder continuous in space,
then density-increments $\delta \rho$ are scale-local. However, the coarse-grained density $\overline{\rho}$
in that case is dominated by energy-scale contributions and infrared-locality breaks down. On the other 
hand, there is evidence from numerical simulations that, for Mach numbers much larger than 1, the density
in compressible fluid turbulence exists only as a distribution (measure) in the infinite Reynolds number limit 
\cite{kim2005density}.  In that case, $\overline{\rho}$ is scale-local but $\delta \rho$ is dominated by 
dissipation-range contributions, and ultraviolet-locality breaks down. In either case, scale-locality
through the dependence on density is always broken in one direction}. 

Similar results hold also for the baropycnal work, where  identical arguments yield a 
``4/5th-law'' type result of the form 
\textcolor{black}{\cite{aluie2011compressible,aluie2013scale,drivas2017onsager}}.
\be  -\grad\bar{p}\bdot\bar{\tau}(\rho,\bv)/\bar{\rho} =O\left(\frac{(\delta p)(\delta \rho)(\delta v)}{\ell\rho}\right). \lb{eq50} \ee
An Onsager singularity theorem for this quantity states that it can be non-vanishing 
as $\ell\rightarrow 0$ only if a condition is satisfied of the form 
\be  \zeta_q^p+\zeta_q^\rho+\zeta_q^v\leq q, \quad \forall q\geq 3 \lb{eq51} \ee
for scaling exponents of all three fields $p,\ \rho,\ \bv$ \textcolor{black}{\cite{aluieunpub,drivas2017onsager}}.
Thus, the baropycnal work contributes
to energy cascade only if the pressure and density, in addition to the velocity, are sufficiently rough. 
When the $q$th-order scaling exponents of all these fields lie between 0 and $q$ then the 
baropycnal work is also a scale-local quantity by the same arguments as for deformation work
\textcolor{black}{\cite{eyink2005locality,aluie2011compressible}}.  
\textcolor{black}{
These inequalities have been stated in terms of singularities for Euler solutions obtained in the limit
of infinite Reynolds numbers, but it is important to emphasize that (\ref{eq49}),(\ref{eq51}) are necessary 
for sustaining an energy cascade rate non-decreasing with $\ell$ at large but finite Reynolds numbers.} 

 \textcolor{black}{The above conclusions on kinetic energy cascade in compressible turbulence
are almost entirely based upon earlier works of  Aluie \cite{aluie2011compressible,aluie2013scale,aluieunpub}, 
and seem to closely parallel the theory of Onsager for incompressible fluids. 
However, we now show that compressibility leads to a novel mechanism for anomalous dissipation of kinetic energy. 
By our fundamental hypothesis (\ref{eq18}), the viscous heating does not vanish at high Reynolds numbers.  
It follows by taking the ideal limit 
$\eta,$ $\zeta,$ $\kappa\rightarrow 0$ of the fine-grained kinetic energy balance (\ref{eq14}) that 
\be  \partial_t(\frac{1}{2}\rho v^2)+\grad\bdot[(p+\frac{1}{2}\rho v^2)\bv]=p*  \Theta-Q_{visc} \lb{eq60} \ee
where $Q_{visc}=Q_\eta+Q_\zeta$ as in (\ref{eq18}) and we have defined 
\be  p*  \Theta = {\mathcal D}\mbox{-}\lim_{\eta,\zeta,\kappa\rightarrow 0}  p\Theta. \lb{eq54} \ee
One might naively conjecture that the latter quantity is the same as $p\circ \Theta$ 
given by (\ref{eq45}). However, the general theory of distributional products makes this 
{\it a priori} highly unlikely. It is part of the definition of the product $f\circ g={\mathcal D}\mbox{-}\lim_{\ell\rightarrow 0}
\bar{f}_\ell \cdot \bar{g}_\ell$ that the 
limiting distribution must be independent of precisely which filter kernel $G$ is employed, but it is generally 
{\it not} true that other regularizations $f_\epsilon,$ $g_\epsilon$ for which $f_\epsilon\Dto f,$ 
$g_\epsilon\Dto g$ will have same limiting product  $f_\epsilon \cdot g_\epsilon\Dto f\circ g$ 
\cite{Oberguggenberger92}.  Since viscosities and thermal conductivities are a different ``regularization''
of the Euler system than mere coarse-graining, one should expect that $p*\Theta\neq p\circ \Theta.$}

\textcolor{black}{Nevertheless, the fine-grained/dissipation-range energy balance (\ref{eq60}) must agree with the 
coarse-grained inertial-range balance (\ref{eq44}) in the limit as $\ell\to 0$. Objective physical 
facts such as the rate of decay of energy or the rate of absorption of power input cannot depend 
upon an arbitrary scale $\ell$ of spatial resolution of observations. After taking first the limit 
$\eta,$ $\zeta,$ $\kappa\rightarrow 0,$ we must then be able to take $\ell\to 0$ and reproduce the same result.  
Comparing (\ref{eq60}) and (\ref{eq44}), it follows necessarily that  
\be  Q_{visc} = \tau(p,\Theta) + Q_{flux}, \lb{eq58} \ee
where we have defined the quantity 
\begin{eqnarray}
\tau(p,\Theta) &=& {\mathcal D}\mbox{-}\lim_{\ell\rightarrow 0} \lim_{\eta,\zeta,\kappa\rightarrow 0} \bar{\tau}(p,\Theta) \cr 
&=& p*  \Theta - p\circ \Theta, 
\lb{eq59} \end{eqnarray} 
which we call the ``pressure-dilatation defect". It is non-vanishing when the joint limits 
$\lim_{\ell\rightarrow 0}$ and $\lim_{\eta,\zeta,\kappa\rightarrow 0}$ of the product  $\bar{p}\cdot \bar{\Theta}$
do not commute, but instead yield either $p*\Theta$ or $p\circ \Theta$
depending upon the order of the two limits. 
Unlike incompressible fluid turbulence where $Q_{visc}=Q_{flux}$ \cite{duchon2000inertial}, we see 
that for compressible fluids the pressure-dilatation defect $\tau(p,\Theta)$ can be another source of 
anomalous dissipation distinct from energy cascade.
In fact, all stationary, planar shocks in fluids with an ideal-gas equation of state exhibit this mechanism in a pure form,
because there $Q_{flux}=0$ and $Q_{visc} = \tau(p,\Theta)\geq 0.$ \\ For a proof of this result, 
see Appendix \ref{Shock}.  All of the anomalous dissipation in such shocks, or so-called ``shock heating'', is due to the 
pressure-dilatation defect. 
In addition to the general inertial-range result that $Q_{flux}=0$ for such shocks, we can also obtain
exact dissipation-range limits at special values of the Prandtl number where analytical results are 
available: $Pr=3/4$ \cite{becker1922stosswelle} and $Pr=0,\ \infty$ \cite{johnson2013analytical}. 
For cases $Pr=3/4,\ \infty$, in particular, we show in  Appendix \ref{Shock} that $\tau(p,\Theta)>0$. }

\textcolor{black}{
Our arguments show generally that $Q_{visc}>0$ only if at least one of $\tau(p,\Theta)$ or $Q_{flux}$ is positive. 
For developed compressible turbulence one should expect that both of these mechanisms will contribute. 
At finite $\ell$ we may rewrite the inertial-range kinetic energy balance (\ref{eq41}) as 
\begin{eqnarray}
&& \partial_t(\frac{1}{2}\bar{\rho} |\tilde{\bv}|^2)+\grad\bdot\Big[(\bar{p}+\frac{1}{2}\bar{\rho} |\tilde{\bv}|^2)\tilde{\bv}
 \cr
&& \qquad +\bar{\rho}\tilde{\tau}(\bv,\bv)\cdot\tilde{\bv}- \frac{\bar{p}}{\bar{\rho}}\bar{\tau}(\rho,\bv) \Big]=\overline{p*\Theta}-Q^{inert}_\ell
\lb{eq41b} \end{eqnarray} 
where 
\be Q^{inert}_\ell=\overline{\tau}_\ell(p,\Theta) + Q^{flux}_\ell \lb{eq61} \ee
is an effective ``inertial dissipation'' at scale $\ell$, such that $Q^{inert}:=\Dlim_{\ell\to 0} Q^{inert}_\ell=Q_{visc}$.  
The effective dissipation at each arbitrary scale $\ell$ can agree with the fine-grained/viscous dissipation rate only if 
there is either nonlinear energy cascade 
\footnote{The notion of ``energy cascade'' has sometimes been criticized as unphysical because it 
depends upon arbitrary scale decompositions. For example, consider the following quotes from 
one prominent critic: 
\begin{quote}
``On the other hand, energy transfer, just like any physical process, should be invariant of particular 
decompositions/representations of a turbulent field.  In this sense Kolmogorov's choice of dissipation 
(and energy input) are well defined and decomposition independent quantities, whereas the energy flux 
is (generally) not, since it is decomposition dependent. After all Nature may and likely does not know 
about our decompositions.'' -- Tsinober \cite{tsinober2009informal} 
\end{quote}
and 
\begin{quote}
``We have seen that there is an ambiguity in defining the meaning of the term `small scales' (or more 
generally `scales' or `eddies', see appendix C) and consequently the meaning of the term `cascade'. ''
-- Tsinober \cite{tsinober2009informal}
\end{quote}
One erroneous statement above is the claim that energy flux is ``generally'' decomposition 
dependent. In fact, the energy cascade rate over a long inertial range at high Reynolds numbers 
is demonstrably the same for any filter kernel satisfying very general, mild assumptions of smoothness and 
rapid spatial decay. However, the other remarks are correct and acute. Indeed,  the physical process 
must be invariant of particular decompositions/representations and independent of the scale 
of observation. What the criticism is missing is that the requirement of such invariance is a positive 
principle which can be exploited to deduce exact consequences.}
with $Q^{flux}_\ell>0$ or a positive defect $\overline{\tau}_\ell(p,\Theta)>0$ as $\ell\to 0.$
The estimates of Aluie \cite{aluie2011compressible,aluie2013scale,aluieunpub} show that the 
fluid variables must be sufficiently rough in order to sustain energy cascade. The consequences 
(\ref{eq49}),(\ref{eq51}) for scaling exponents are directly testable predictions of the argument, which 
is an exact, non-perturbative application of the  principle of renormalization-group invariance 
\cite{stueckelberg1951normalization,gellmann1954quantum,bogolyubov1955group}. Although 
it is not yet obvious, the condition that $\overline{\tau}_\ell(p,\Theta)>0$ as $\ell\to 0$ also requires these 
same exponent relations to hold. In order show this, we must develop a deeper understanding of the 
thermodynamics of compressible turbulence.} 

\subsection{Internal Energy}\label{sec:internal}

\textcolor{black}{
The other half of the energy budget is internal energy. Numerical results \cite{aluie2012conservative,ni2015effects}
show that up to 50\% of the energy injected at large scales can be channeled into internal energy by the 
large-scale pressure work. We must therefore consider the inertial-range dynamics of internal energy. 
The simplest way to obtain an equation for the coarse-grained/resolved internal energy $\bar{u}$ is to apply
the coarse-graining operation to the equation (\ref{eq15}) for fine-grained internal energy and then to consider
the ideal (infinite Reynolds and P\'eclet number) limit. The first step yields
\be  \partial_t \overline{u}+\grad\bdot\left[\overline{u\bv-\kappa\grad T}\right]=-\overline{p\Theta}
+\overline{2\eta|\bS|^2+\zeta \Theta^2}. \lb{eq52} \ee
Invoking the fundamental hypothesis (\ref{eq18}) and taking the ideal limit $\eta,$ $\zeta,$ $\kappa\rightarrow 0$
yields the following equation for the inertial-range dynamics of the internal energy: 
\be  \partial_t \bar{u}+\grad\bdot(\bar{u}\bar{\bv}+\bar{\tau}(u,\bv))=-\overline{p*  \Theta}+\bar{Q}_{visc}. \lb{eq55} \ee
In the subsequent limit $\ell\rightarrow 0$ we get 
\be  \partial_t u+\grad\bdot(u\bv)=-p* \Theta+Q_{visc}. \lb{eq56} \ee
as the distributional balance of internal energy for the limiting weak Euler solution.}

\textcolor{black}{
On the other hand, we can obtain another form of this equation by subtracting  the balance equation (\ref{eq41}) for 
resolved kinetic energy from the coarse-graining of equation (\ref{eq37}) for conservation of total energy. 
This yields after some straightforward calculations the equation 
\begin{eqnarray}  
&& \partial_t \left(\overline{u}+\frac{1}{2}\overline{\rho}\tilde{\tau}(v_i,v_i)\right)
+\grad\bdot\Big(\overline{u}\, \overline{\bv}+\overline{\tau}(h,\bv) + \frac{1}{2}\overline{\rho}\tilde{\tau}(v_i,v_i)\tilde{\bv} \cr
&& \hspace{50pt} +\frac{1}{2}\overline{\rho}\tilde{\tau}(v_i,v_i,\bv)\Big)=
-\overline{p}\overline{\Theta}+Q^{flux}_\ell. \lb{eq57} 
\end{eqnarray}
with $h=u+p$ the enthalpy. There appears in this balance the quantity 
\be \overline{u}^*:=\overline{u}+\frac{1}{2}\overline{\rho}\tilde{\tau}(v_i,v_i), \lb{eq57a} \ee
which we shall call the {\it intrinsic large-scale/resolved internal energy.} It is a natural object because,
based on coarse-grained observations alone, it is impossible to distinguish between energy in thermal
fluctuations and in unresolved turbulent fluctuations. In contrast to the balance (\ref{eq55}) for 
resolved internal energy, the balance (\ref{eq57}) for intrinsic large-scale internal energy contains 
no direct contributions from microscopic dissipation and is a consequence solely of the limiting 
distributional Euler solution. In the limit as $\ell\to 0$ all of the cumulant terms in (\ref{eq57}) 
vanish distributionally and one obtains a second form of the internal energy balance:   
\be \partial_t u+\grad\bdot(u\bv)=-p\circ \Theta+Q_{flux}. \lb{eq57b} \ee
Using (\ref{eq61}), we may rewrite the righthand side of (\ref{eq57b}) as $-p*\Theta+Q_{inert}$. 
The two equations (\ref{eq56}) and (\ref{eq57b}) thus agree, 
since $Q_{visc}=Q_{inert}=Q.$ We see that the same $Q$ which appears as a sink in the kinetic 
energy balance, (\ref{eq60}) or (\ref{eq41b}), appears as a source in the balance of internal energy, 
(\ref{eq56}) or (\ref{eq57b}).} 

One concern at this point is whether the equation (\ref{eq55}) for $\bar{u}$ truly represents ``inertial-range
dynamics,'' in contrast to equation (\ref{eq57}) for $\overline{u}^*$ which is clearly an inertial-range balance. 
Quantities of the sort $\bar{Q}_{visc}$ have been much discussed 
for incompressible fluid turbulence in the context of the ``Kolmogorov refined simlarity hypothesis''
\cite{kolmogorov1962refinement} and it has been a debated issue whether such coarse-grained dissipation fields 
for inertial-range lengths $\ell$ should be considered inertial-range or dissipation-range. For example, Kraichnan
\cite{kraichnan1974kolmogorov}
concluded that $\bar{Q}_{visc}$ is not inertial-range and ``Instead it is the integral of a dissipation-range quantity.''   
The question is hard to argue substantively, because there is no clear, accepted definition in the literature of 
what it means to be ``inertial-range'' or ``dissipation-range''. We would like to offer a simple, precise definition of an 
``inertial-range quantity'' as any field which exists as an ordinary function (as opposed to a distribution only)
in the ideal limit $\eta,$ $\zeta,$ $\kappa\rightarrow 0.$ By this definition, $\bar{Q}_{visc}$ is clearly an 
inertial-range quantity and so is the pressure work $\overline{p*\Theta},$ although both 
involve effects of molecular dissipation which survive in the ideal limit $\eta,$ $\zeta,$ $\kappa\rightarrow 0.$  

The questions about inertial-range status of $\overline{Q}_{visc}$ and $\overline{p*\Theta}$ cannot however 
be legitimately answered by merely offering a definition. The more serious worry which underlies 
this question is whether these can be universal quantities independent of the particular 
micro-scale dissipation mechanism, or whether they shall be distinct for every particular 
fine-grained dissipation (e.g. ordinary viscosity vs. hyper-viscosity). As a matter of fact,
the quantities $\overline{Q}_{visc}$ and $\overline{p*\Theta}$ probably cannot be completely 
universal in compressible fluid turbulence, as they can be shown to be Prandtl-number dependent 
\footnote{At least the results are distinct for $Pr=0,$ $Pr=\infty,$ and $0<Pr<\infty.$ 
We have analytical results for only one finite positive value $Pr=3/4$}
even for planar, ideal-gas shocks (see Appendix A). On the other hand, it is a direct consequence of 
equation (\ref{eq56}) for the internal energy $u$ that the combination $-p*\Theta+Q_{visc}$ 
depends only upon the limiting Euler solution fields $\rho,$ $\bv,$ $u$ and not upon the 
particular sequence $\eta,$ $\zeta,$ $\kappa\rightarrow 0$ used to obtain that solution. 
It is explicitly verified for planar, ideal-gas shocks in Appendix A that the combined quantity 
is independent of Prandtl-number even though the quantities $\overline{Q}_{visc}$ and $\overline{p*\Theta}$ 
separately are $Pr$-dependent. This suggests that the combination $\overline{Q}_{visc}-\overline{p*\Theta}$ 
for inertial-range length-scales $\ell$ should be universal for a wide class of fine-grained dissipation 
mechanisms and determined only by fluid modes at scales comparable to $\ell$. 

We can make a substantive argument for this assertion based upon the following equation for 
subscale/unresolved kinetic energy in the ideal limit: 
\begin{eqnarray}
&& \partial_t \left(\frac{1}{2}\bar{\rho}\tilde{\tau}(v_i,v_i)\right)
+\grad\bdot\left[\frac{1}{2}\bar{\rho}\tilde{\tau}(v_i,v_i)\tilde{\bv} +\bar{\tau}(p,\bv) \right. \cr
&& \qquad \qquad \qquad \qquad \qquad \qquad \quad + \left. \frac{1}{2}\bar{\rho}\tilde{\tau}(v_i,v_i,\bv)\right] \cr
&& \qquad \qquad \qquad = \bar{\tau}(p,\Theta) - \bar{Q}_{visc} + Q_\ell^{flux}. 
\lb{eq62} \end{eqnarray} 
This equation is straightforward to derive by considering the equations for $(1/2)\overline{\rho |\bv|^2}$
and $(1/2)\bar{\rho} |\tilde{\bv}|^2,$ subtracting them, and taking the limit $\eta,$ $\zeta,$ $\kappa\rightarrow 0.$ 
A simple re-organization of this equation gives
\begin{eqnarray}
&& \overline{p*\Theta} - \bar{Q}_{visc}= \bar{p}\bar{\Theta} - \bar{Q}_{flux} \cr
&& + \frac{1}{2}\bar{\rho}\tilde{D}_t\tilde{\tau}(v_i,v_i) + \grad\bdot\left[\bar{\tau}(p,\bv) 
+ \frac{1}{2}\bar{\rho}\tilde{\tau}(v_i,v_i,\bv)\right],
\lb{eq63} \end{eqnarray}
where $\tilde{D}_t=\partial_t+\tilde{\bv}\bdot\grad$ is the large-scale Lagrangian time-derivative. 
The important point is that all of the terms on the right-hand side of this equation are pure 
inertial-range quantities that are local-in-scale and thus determined only by fluid modes near 
the considered scale $\ell.$ The standard arguments for universality thus apply to the 
righthand side and so we may argue that as well the lefthand side, the combination $\overline{p*\Theta} - 
\overline{Q}_{visc},$ will be a universal, inertial-range quantity, independent of the particular microscale 
mechanism of dissipation.  As we shall discuss further in the following section, the above considerations 
play an essential role in our proof of a complete Onsager theorem for compressible turbulence 
\cite{drivas2017onsager}. \textcolor{black}{By this we mean the proof that a kinetic-energy dissipation anomaly $Q\neq 0$ 
requires singularities in the fluid fields. As emphasized earlier, the arguments of Aluie 
\cite{aluie2011compressible,aluie2013scale,aluieunpub} imply that $Q_{flux}\neq 0$ requires the inequalities 
(\ref{eq49}),(\ref{eq51}) to hold, but it is still possible in principle that $Q_{visc}=\tau(p,\Theta)>0$ with milder 
singularities.}  

Let us close this section by briefly considering the energy balances that must exist in a long-time
steady-state of mechanically forced compressible turbulence. In order for a steady state to exist one must 
take into account cooling mechanisms, such as electromagnetic radiation, otherwise the internal energy 
will continue to grow due to input from viscous dissipation and mechanical work. The situation may be 
modelled by the compressible Navier-Stokes equations modified to include an external acceleration field ${\bf a}_{ext}$
and a cooling function $Q_{cool}$:
\be \partial_t\rho+\grad\bdot(\rho\bv)=0, \lb{eq64} \ee
\be  \partial_t(\rho\bv)+\grad\bdot(\rho\bv\bv+p\bI-2\eta\bS-\zeta \Theta\bI)=\rho {\bf a}_{ext}, \lb{eq65} \ee
\begin{eqnarray}
&&  \partial_t(\frac{1}{2}\rho v^2+u)+\grad\bdot\big[(u+p+\frac{1}{2}\rho v^2)\bv \cr 
&& \hspace{20pt} -\kappa \grad T-2\eta \bS\bdot\bv-\zeta \Theta\bv\big] \ =\  \rho \bv\bdot {\bf a}_{ext} - Q_{cool}. 
\lb{eq66} \end{eqnarray}
The acceleration field is a source of mechanical input of kinetic energy $Q_{in}=\rho \bv\bdot {\bf a}_{ext},$
whereas the equation (\ref{eq15}) for internal energy now includes the cooling term $-Q_{cool}$ 
on the right. When the forcing ${\bf a}_{ext}$ and cooling function $Q_{cool}$ are large-scale (smooth) fields,
then all of our previous considerations on the ideal limit apply. Steady-state kinetic energy balance 
gives $\langle Q_{in}\rangle =\langle Q_{trans}\rangle +\langle Q\rangle$, where $Q_{trans}=-p* \Theta.$ 
From the fine-grained point of view, $Q=Q_{visc}$ whereas in the inertial-range $Q=Q_{inert}$ and 
$\langle Q_{inert}\rangle=\langle Q_{in}\rangle -\langle Q_{trans}\rangle,$ exactly as argued earlier by Aluie 
\cite{aluie2013scale}.
The steady-state internal energy balance likewise gives $\langle Q_{trans}\rangle+\langle Q\rangle =
\langle Q_{cool}\rangle,$ so that $\langle Q_{in}\rangle=\langle Q_{cool\rangle}$ in the steady-state.
In decaying turbulence without external forcing such as we considered throughout 
most of the paper, one expects a quasi-steady state with initial conditions supplying the reservoir of energy 
to drive the cascade and $-\langle \partial_t (\rho v^2/2)\rangle$ playing the role of $\langle Q_{in}\rangle.$
Likewise, if cooling mechanisms are inefficient, then $\langle \partial_t u\rangle $ plays the role of $\langle Q_{cool}\rangle$.

\section{Entropy Cascade}\label{sec:entropy} 

The energy transfer $\bar{p}\bar{\Theta}$ from large-scale kinetic energy to large-scale 
internal energy $\bar{u}$ which was discussed in the previous section does not represent a global 
heating of the fluid resulting merely in a uniform increase in the internal energy. It is instead an ``ordered'' or 
``coherent'' input of energy, which leads to large-scale structure of the internal energy field $\bar{u}$. 
One should thus expect this input to decrease the large-scale entropy of the system, which is maximum 
for a spatially homogeneous state. These considerations are one motivation to consider in detail the entropy balance
of the turbulent flow, which allows us to verify the above expectations. Because of the constraints imposed 
by the second law of thermodynamics, the entropy in fact turns out to play a central role in the entire theory 
of compressible fluid turbulence. 

We recall that the entropy per volume $s(u,n)$ is a concave function of the internal energy per 
volume $u$ and the particle number per volume $n.$ This follows microscopically from the  
extensivity of the thermodynamic limit 
\cite{ruelle1999statistical,martin1979statistical} and macroscopically from 
the stability of thermodynamic equilibrium \cite{callen1985thermodynamics}.  The entropy is also an analytic function,
except at phase transitions, and we restrict our discussion here to single-phase flows.  
The quantity $s(\bar{u},\bar{n})$ naturally
represents the ``large-scale/resolved entropy'' for a given length-scale $\ell.$ We can in turn define 
the ``small-scale/unresolved entropy''
\be  \Delta s\equiv \overline{s(u,n)}-s(\bar{u},\bar{n})\leq 0. \lb{eq67} \ee
The non-positivity follows by concavity,  so that spatial coarse-graining increases entropy.   
We use throughout our 
discussion below the shorthand notations 
\be \overline{s}=\overline{s(u,n)}, \quad \underline{s}=s(\bar{u},\bar{n}), \lb{eq68} \ee
and likewise for other thermodynamic functions of $u,n.$ Note that $\langle \us\rangle$ 
plays the role of a ``cumulative entropy (co)spectrum'' up to wave-number $\sim 1/\ell$ and 
$\langle \Delta s\rangle$ is analogous to a second-order ``entropy structure function'' at separation $\ell.$
In many respects it is more natural to consider a quantity $s_{\max}-s$ which is convex and 
decreasing in time rather than the traditional entropy, which is instead concave and increasing. The 
quantity $s_{\max}-s$  was known variously as ``capacity of entropy'' by 
Gibbs \cite{gibbs1873method}, ``deficiency of entropy'' by Obukhov \cite{obukhov1949temperature}, 
and ``negentropy" by Brillouin \cite{brillouin1953negentropy}.  We use here the latter term. 

We first derive the balance equation of large-scale entropy at finite $\eta,\ \zeta,\ \kappa$
using the equation (\ref{eq55}) for $\bar{u}$ and (\ref{eq27}) for $\bar{\rho},$ with 
$\overline{\rho\bv}=\bar{\rho}\,\bar{\bv}+\bar{\tau}(\rho,\bv).$  
Invoking the first law of thermodynamics in the form $du=Tds+\mu dn$ 
for absolute temperature $T$ and chemical potential $\mu,$
and denoting $\bar{D}_t= \partial_t + \bar{\bv}\bdot \grad,$ we get from 
\be  \uT \bar{D}_t \us = \bar{D}_t \bar{u} - \umu \bar{D}_t\bar{n}\lb{eq69} \ee 
after some straightforward calculation that 
\be 
\partial_t\us+\grad\bdot [\us\bar{\bv} +\ubeta\ \bar{\tau}(u,\bv) -\ulambda\ \bar{\tau}(n,\bv)]  = 
 \Sigma_\ell^{inert}, \lb{eq70} \ee
with inertial-range entropy production given by 
\be \Sigma_\ell^{inert} = -I_\ell^{mech} +\Sigma_\ell^{flux} + \bar{Q}_{visc}/\uT, \lb{eq72} \ee
 for mechanical input of negentropy 
\be  I_\ell^{mech} = \frac{\overline{p\Theta}-\up\overline{\Theta}}{\uT} \lb{eq71} \ee
and for negentropy flux 
\be  \Sigma_\ell^{flux}=\grad\ubeta\bdot \bar{\tau}(u,\bv)-\grad\ulambda \bdot \bar{\tau}(n,\bv). \lb{eq73} \ee
The quantity $\beta=1/T$ in the expressions above is inverse temperature and $\lambda=\mu/T$
is the thermodynamic potential entropically conjugate to particle number.
We now discuss in detail the physical significance of each of these various contributions 
to the entropy balance. 

First, $I_\ell^{mech}$ represents the net input of negentropy \footnote{Here the letter ``I''
may stand either for ``input'', or for ``information'', which is another synonym for negentropy.} 
into the large-scales from pressure work,
where $-\up\bar{\Theta}/\uT$ is the ``coherent input'' of negentropy at large-scales and 
$\overline{p\Theta}/\uT$ is the entropy production (destruction of negentropy) due to 
mechanical heating at all scales. There is competition between these two terms
but, as anticipated, they will not cancel in general. We suggest that it is likely that 
$\langle \up\bar{\Theta}/\uT\rangle<\langle \overline{p\Theta}/\uT\rangle<0,$ because of the greater 
coherence at larger scales and the near cancellations between positive and negative terms 
at small scales \cite{aluie2012conservative,ni2015effects}.  
An alternative decomposition to that above is 
\be  I^{mech}_\ell= \frac{\bar{\tau}(p,\Theta) }{\uT}+I^{flux}_\ell, \quad I^{flux}_\ell=\frac{\Delta p \ \bar{\Theta}}{\uT}\lb{eq74} \ee
with $\Delta p=\bar{p}-\up.$ The first term is related to the pressure-dilatation defect and the second term 
is  a ``flux-like'' contribution, in the sense that it represents an interaction between a large-scale dilatation
$\bar{\Theta}$ and a small-scale pressure $\Delta p$. There is a simple formula for the latter 
\cite{drivas2017onsager} \footnote{For completeness, we note here without proof the 
relevant identity. Let $f(\brho)$ be any smooth function of a set of density functions  
$\brho=(\rho_1,...,\rho_m)$ and let $\brho(\bx)$ be a spatial field of the densities. Then,
$\triangle f=\overline{f(\rho)}-f(\overline{\rho})$ is equal to 
$\int d^dr\ G_\ell(\br) [f(\brho(\bx+\br))-f(\brho(\bx))-\delta \brho(\bx;\br)\bdot(\grad f)(\brho(\bx))] $
minus $f(\overline{\brho}(\bx))-f(\brho(\bx)) + \brho^{\prime}(\bx)\bdot (\grad f)(\brho(\bx)),$
defining $\delta \brho(\bx;\br)=\brho(\bx+\br)-\brho(\bx),$ $\brho'(\bx)=\brho(\bx)-\overline{\brho}(\bx)$
} which provides an estimate
\be \Delta p = O\left((\delta u)^2,(\delta u)(\delta \rho),(\delta \rho)^2\right)\lb{eq75} \ee
yielding a ``4/5th-law''
type representation of the flux term. It is worth noting that for an ideal gas with adiabatic index $\gamma=c_P/c_V,$ 
one has $p(u,n)=(\gamma-1) u$ so that $\Delta p=0$ exactly and the flux-like term is absent. 
Neither term is present in a naive fine-grained calculation.

Secondly, $\Sigma_\ell^{flux}$ represents negentropy flux to small scales, arising from small-scale 
turbulent transport of heat energy $\bar{\tau}(u,\bv)$ acting against large-scale temperature gradients $\grad\uT$ 
and small-scale particle transport $\bar{\tau}(n,\bv)$ acting against large-scale $\ulambda$-gradients. These 
contributions will be positive, indicating entropy production/inverse cascade of entropy/forward cascade of negentropy, 
when the turbulent transport is ``down-gradient'', with heat-transport from higher to lower resolved temperatures $\ubeta$
and particle-transport from higher to lower $\ulambda$-potential. Because of lack of scale-separation of 
turbulent transport both positive and negative values will occur pointwise in space-time for finite $\ell$,
but one should expect that 
on average $\langle \Sigma_\ell^{flux}\rangle>0,$ consistent with the overall increase of entropy from 
the second law of thermodynamics.
Of course, the term $\bar{Q}_{visc}/\uT\geq 0$ is the resolved entropy production at large scales due to 
viscous dissipation.


\textcolor{black}{
We now consider the situation when there is anomalous entropy production in the 
ideal limit $\eta,\ \zeta,\ \kappa\rightarrow 0,$ as hypothesized in (\ref{eq20}). The same 
result must be obtained by considering either the fine-grained entropy balance (\ref{eq19})
or the inertial-range balance (\ref{eq70}). Indeed, because of concavity of the volumetric entropy density, 
the total entropy observed ``without spectacles'' at resolution $\ell$ can only exceed the true entropy 
\be
 S_\ell(t)= \int d^dx  \ s(\bar{u}_\ell,\bar{\rho}_\ell) 
  \geq \int d^dx  \ s(u,\rho)=S(t) \lb{entropycomp} \ee 
and in the limit $\ell\to 0$ they must agree. Thus, if entropy $S(t)$ continues to grow 
over a finite time-interval in the limit $\eta,\ \zeta,\ \kappa\rightarrow 0,$ then $S_\ell(t)$ 
must also grow for the subsequent limit $\ell\to 0$ described by a weak Euler solution. 
Taking the limit $\eta,\ \zeta,\ \kappa\rightarrow 0$ of the fine-grained entropy balance (\ref{eq19}), 
the anomaly is represented as 
\be  \partial_t s+\grad\bdot(s\bv)= \Sigma^{diss} \lb{eq79} \ee
with $\Sigma^{diss}=\Sigma^{therm}+\Sigma^{visc},$
for viscous entropy production 
\be  \Sigma^{visc} = \beta*  Q_{visc} = {\mathcal D}\mbox{-}\lim_{\eta,\zeta,\kappa\rightarrow 0} \beta Q_{visc} \lb{eq80} \ee
and for entropy production by thermal conduction
\be  \Sigma^{therm} = {\mathcal D}\mbox{-}\lim_{\eta,\zeta,\kappa\rightarrow 0} \frac{\kappa|\grad T|^2}{T^2}. \lb{eq81} \ee
The coarse-grained entropy balance (\ref{eq70}) in the limit $\eta,\ \zeta,\ \kappa\rightarrow 0$ 
is unchanged, except that $\overline{p\Theta}\rightarrow \overline{p*\Theta}.$ In the subsequent limit 
$\ell\rightarrow 0$ the inertial-range entropy balance becomes  
\be  \partial_t s+\grad\bdot(s\bv)= \Sigma^{inert}, \lb{eq76} \ee
where  $\Sigma^{inert}=-I^{mech}+\Sigma^{flux}+\beta\circ Q_{visc}$ with \footnote{We are not being entirely consistent
with our mathematical notations. If so, we should have defined $\beta\circ Q_{visc} = \lim_{\ell\rightarrow 0} 
\overline{\beta}\cdot \overline{Q}_{visc}$ and the quantities in equations (\ref{eq78}) and (\ref{eq77}) would require a new 
notation. To avoid a proliferation of new symbols, we use the same notation ``$\circ$'' everywhere for the 
relevant distributional products obtained by limits of $\ell$ through the inertial-range, whereas the notation
``$*$'' stands for the distributional product obtained by limits $\eta,\ \zeta,\ \kappa\rightarrow 0$ of 
fine-grained/disssipation-range quantities.}
\be  \beta\circ Q_{visc} = {\mathcal D}\mbox{-}\lim_{\ell\rightarrow 0} \ubeta \bar{Q}_{visc}.\lb{eq78} \ee
and where $I^{mech}=I_{flux}+\beta\circ \tau(p,\Theta)$ with 
\be  \beta\circ \tau(p,\Theta) = {\mathcal D}\mbox{-}\lim_{\ell\rightarrow 0} \ubeta \bar{\tau}(p,\Theta). \lb{eq77} \ee
The expressions (\ref{eq76})-(\ref{eq77}) provide an inertial-range representation of anomalous entropy
production. Equating the two different expressions, $\Sigma_{inert}=\Sigma_{diss}.$}

The above general results are nicely illustrated by planar shocks in an ideal gas for the special value 
of Prandtl number $Pr=3/4$.  There is in fact a 2-parameter family of stationary shocks in ideal gases, 
labelled by the adiabatic index $\gamma>1$ and by the pre-shock Mach number $M_0>1,$ 
or, alternatively, the compression ratio $R=(\gamma+1)/[(\gamma-1)+2/M_0^2].$ All of the anomalous quantities in the balances 
above are non-zero for $Pr=3/4$ (with the exception of $I_{flux}$, since $\Delta p\equiv 0$  
for an ideal gas) and are proportional to Dirac delta functions at the location of the shock. 
For a shock situated at the origin there is a positive entropy production anomaly 
of the very simple form
\be \Sigma_{inert}=\Sigma_{diss}=(\Delta s_m)j_*\delta(x), \lb{eq83} \ee
where $\Delta s_m=s_{m,1}-s_{m,0}>0$ is the jump across the shock of the entropy per mass $s_m=s/\rho$
(with ``0'' denoting the gas in front of the shock and ``1'' gas behind the shock) and $j_*=j_0=j_1>0$ is the 
mass flux through the shock. Explicit expressions for all terms in the entropy balance are given in Appendix \ref{Shock}. 
As expected, the inertial-range quantities $\Sigma_{flux}$ and $\beta\circ (Q_{visc}-\tau(p,\Theta))$
in the infinite Reynolds-number limit are identical for all planar, ideal-gas shocks with the same values of $\gamma$ and $M_0,$
entirely independent of the precise molecular dissipation mechanism. In particular, these two particular 
quantities are Prandtl-number independent.  It is interesting that the negentropy flux is non-zero for such shocks, even though the 
energy flux vanishes exactly. This shows clearly that the two cascades are distinct in general. In fact, it is the negentropy 
flux $\Sigma_{flux}$ which inside the inertial-range of planar, ideal-gas shocks supplies the contribution $(\Delta s_m)j_*\delta(x),$ 
arising from the particle transport term $-\grad\ulambda \bdot \bar{\tau}(n,\bv).$ All other inertial-range contributions 
cancel between $I_{mech},$ $\Sigma_{flux},$ and $\beta\circ Q_{visc}.$  
See Appendix \ref{Shock} for details. 

While Euler shock solutions with discontinuous fields provide a simple example where the negentropy 
flux is non-vanishing, more modest singularities are able to support a negentropy cascade. 
Note from the formula (\ref{eq75}) for $\Delta p$ and the definition (\ref{eq74}) of $I^{flux}_\ell$ that 
\be I^{flux}_\ell= O\left((\delta u)^2,(\delta u)(\delta \rho),(\delta \rho)^2\right) O\left(\frac{\delta v}{\ell}\right). \lb{eq84} \ee
Hence, this term may have a non-vanishing limit as $\ell\rightarrow 0$ whenever 
\be  2\min\{\zeta_q^u,\zeta_q^\rho\}+\zeta_q^v\leq q, \quad q\geq 3. \lb{eq85} \ee 
Likewise, from the chain rule for gradients of the smooth functions $\beta(u,\rho)$ and $\lambda(u,\rho)$ 
one gets 
\be  \grad\ubeta,\ \grad\ulambda = \frac{O\left(\delta u,\delta \rho\right)}{\ell} \lb{eq86} \ee
which, with the general result $\tau(f,g)=O(\delta f\cdot\delta g)$ and the definition 
of $\Sigma^{flux}_\ell,$ gives the identical estimate
\be \Sigma^{flux}_\ell= O\left((\delta u)^2,(\delta u)(\delta \rho),(\delta \rho)^2\right) O\left(\frac{\delta v}{\ell}\right). \lb{eq87} \ee
Thus, the inequality (\ref{eq85}) again provides a necessary condition for a non-vanishing limit as $\ell\rightarrow 0.$
The shock solutions with discontinuous fields have $\zeta_q^u=\zeta_q^\rho=\zeta_q^v=1$ for $q\geq 1$
and thus satisfy these inequalities for all $q\geq 3.$ However, multifractal fields $u,$ $\rho,$ $\bv$ with 
positive H\"older exponents can also easily satisfy these inequalities \cite{uriel1995turbulence}. Thus,
for compressible turbulent flow the anomalous entropy production should generally arise not just 
from shocks with zero H\"older exponents but from the spectrum of milder H\"older singularities. 

When the singularity conditions (\ref{eq49}),(\ref{eq51}) are not satisfied, then one expects that 
entropy will in fact be conserved. This statement is an analogue of the Onsager singularity theorem for 
a dissipative anomaly of negentropy. Such a result does not follow directly from the estimates (\ref{eq84}),(\ref{eq87}) 
on the fluxes,  because of the additional terms contributing to the inertial-range entropy balance. 
However, such a result may be proved \cite{drivas2017onsager}, by the following arguments. 
First, rewrite the inertial-range entropy balance as 
\begin{eqnarray}
&& \partial_t\us+\grad\bdot [\us\bar{\bv} +\ubeta\ \bar{\tau}(u,\bv) -\ulambda\ \bar{\tau}(n,\bv)]  = \cr
&& \qquad \Sigma_\ell^{flux}-I^{flux}_\ell +\ubeta \left(\bar{Q}_{visc}-\bar{\tau}(p,\Theta)\right)
\lb{eq88} \end{eqnarray} 
The first two terms on the right are those which have been shown to vanish as $\ell\rightarrow 0$
when (\ref{eq85}) is not satisfied. To evaluate the last term, we use the equation (\ref{eq62}) for the subscale 
kinetic energy. The first two terms on the right of (\ref{eq62}) are exactly those appearing in the entropy balance, 
while the third is the energy flux.  Thus, multiplying (\ref{eq62}) by $\ubeta$ gives 
\begin{eqnarray} 
&& \ubeta\left(\bar{Q}_{visc}-\bar{\tau}(p,\Theta)\right) = \ubeta Q_\ell^{flux} + (\partial_t \ubeta) \frac{1}{2}\bar{\rho}\tilde{\tau}(v_i,v_i) \cr
&& + \grad\ubeta\bdot\left[\frac{1}{2}\bar{\rho}\tilde{\tau}(v_i,v_i)\tilde{\bv} +\bar{\tau}(p,\bv) + \frac{1}{2}\bar{\rho}\tilde{\tau}(v_i,v_i,\bv)\right] 
+ (\cdots)  \cr
&& \lb{eq89} \end{eqnarray}
where $(\cdots)$ denotes a total derivative term which vanishes distributionally in the limit $\ell\rightarrow 0.$
The other three terms are all ``flux-like''. The first of these contains a time-derivative, which is perhaps unexpected, 
but the physical meaning is clearly an entropy-production due to rate of change of large-scale
inverse-temperature times subscale kinetic-energy. The term $\ubeta Q_\ell^{flux}$ is an entropy-production
due to kinetic energy cascade and the terms proportional to $\grad\ubeta$ are a corrrection to the turbulent
internal energy transport $\bar{\tau}(u,\bv).$ These flux terms vanish as explicit power-laws in the limit $\ell\rightarrow 0$
for solutions that are not sufficiently singular. Precisely, at least one of the following conditions 
must be satisfied 
\begin{align}
\lb{eq89-1}
2\min\{\zeta^u_q,\zeta^\rho_q\}+\zeta_q^v &\leq q \qquad q\geq 3 \\
\lb{eq89-2}
\min\{\zeta^u_q,\zeta^\rho_q\}+2\zeta_q^v &\leq q \qquad q\geq 3 \\
\lb{eq89-3}
3\zeta_q^v &\leq q \qquad q\geq 3, 
\end{align}
if the fluxes are not to vanish. Here (\ref{eq89-1}) is the same as (\ref{eq85}), and 
(\ref{eq51}) has also been replaced by (\ref{eq89-1}), which implies vanishing 
of baropycnal work via (\ref{eq50}). 

There is, in fact, a much more fundamental way to reach the same conclusion. Let us define an {\it intrinsic 
large-scale/resolved entropy density} by 
\be \us^*=\us+ \frac{1}{2}{{\underline{\beta}}}\overline{\rho}\ \widetilde{\tau}(v_i,v_i). \lb{eq89a} \ee 
From the homogeneous Gibbs relation $\us={\underline{\beta}}(\overline{u}+{{\underline{p}}})-{\underline{\lambda}}\overline{n},$
it follows that $\us^*={\underline{\beta}}(\overline{u}^*+{{\underline{p}}})-{\underline{\lambda}}\overline{n},$ 
where $\overline{u}^*$ is the ``intrinsic large-scale internal energy'' that was introduced in (\ref{eq57a}). 
Using the equation (\ref{eq57}) for the intrinsic internal energy, the coarse-grained mass conservation 
equation (\ref{eq27}), and the standard thermodynamic relation $d(\underline{\beta}\,\underline{p})=
\overline{n} d\underline{\lambda}-\overline{u} d\underline{\beta},$ it is then straightforward to verify the 
entropy balance equation 
\be \partial_t\us^* + \grad\bdot {\underline{{\bf s}}}^* = \Sigma_\ell^{inert*} \lb{eq89b} \ee 
where 
\begin{eqnarray}
&& {\underline{{\bf s}}}^*=\us\overline{\bv}+{\underline{\beta}}{\overline{\tau}}(u,\bv)-{\underline{\lambda}}{\overline{\tau}}(n,\bv)\cr
&& +{\underline{\beta}}\left[\frac{1}{2}\overline{\rho}\ \widetilde{\tau}(v_i,v_i)\widetilde{\bv}
+\frac{1}{2}\overline{\rho}\ \widetilde{\tau}(v_i,v_i,\bv)+\overline{\tau}(p,\bv)\right], \cr 
&& \lb{eq89c} 
\end{eqnarray} 
is the spatial-current of intrinsic entropy and where $\Sigma_\ell^{inert*} =-I_\ell^{flux}+\Sigma_\ell^{flux*},$ with
\begin{eqnarray}
&& \Sigma_\ell^{flux *} = \Sigma_\ell^{flux} + \underline{\beta} Q_\ell^{flux} + (\partial_t \ubeta) \frac{1}{2}\bar{\rho}\tilde{\tau}(v_i,v_i) \cr 
&& + \grad\ubeta\bdot\left[\frac{1}{2}\bar{\rho}\tilde{\tau}(v_i,v_i)\tilde{\bv} + \frac{1}{2}\bar{\rho}\tilde{\tau}(v_i,v_i,\bv)+\bar{\tau}(p,\bv) \right] \cr
&&\cr
&&  = \grad\ubeta\bdot\left[ \bar{\tau}(h,\bv)+\frac{1}{2}\bar{\rho}\tilde{\tau}(v_i,v_i)\tilde{\bv} + \frac{1}{2}\bar{\rho}\tilde{\tau}(v_i,v_i,\bv)\right] \cr
&&  \, \vspace{40pt} -\grad\ulambda \bdot \bar{\tau}(n,\bv) +  \underline{\beta} Q_\ell^{flux}+ (\partial_t \ubeta) \frac{1}{2}\bar{\rho}\tilde{\tau}(v_i,v_i) 
\cr
&& \lb{eq89d} \end{eqnarray} 
the flux of intrinsic inertial-range negentropy. Although this result is the same as that obtained by substituting (\ref{eq89}) into 
(\ref{eq88}), the present derivation is more general, because it makes no reference to any microscopic model. Thus, 
the equation (\ref{eq89b}) is seen to be valid for all distributional Euler solutions, including those derived from Boltzmann
kinetic theory \cite{yu2005hydrodynamic} for example and not restricted to limits of compressible Navier-Stokes solutions.   

Taking the limit as $\ell\to 0$ of the inertial-range balance equation (\ref{eq89b}) yields again the 
limiting balance (\ref{eq76}) for the  distributional solution of the compressible Euler equations. One concludes 
that any solution that is too regular, 
satisfying none of the conditions (\ref{eq89-1})-(\ref{eq89-3}), will obey local entropy conservation:  
\be  \partial_t s+\grad\bdot(s\bv)= 0. \lb{eq90} \ee
Put another way, $\Sigma_{diss}=0$ unless the fluid fields possess 
singularities compatible with (\ref{eq89-1})-(\ref{eq89-3}). For details of the proof, see the companion paper 
\cite{drivas2017onsager}.  
These arguments also allow one to complete the proof that energy dissipation anomalies
must vanish when (\ref{eq89-1})-(\ref{eq89-3}) are not satisfied.  Because of the non-negativity of the separate viscous and 
thermal conduction contributions to anomalous entropy production, $\Sigma_{diss}=0$
immediately implies that 
\be  \Sigma_{therm}=0, \quad \Sigma_{visc}=\beta*  Q_{visc}=0. \lb{eq91} \ee
It is not hard to see that the second condition implies that $Q_{visc}=0,$ since $\beta\circ Q_{visc}\geq Q_{visc}/\|T\|_\infty\geq 0$ 
\cite{drivas2017onsager}. Coupled with $Q_{flux}=0$ we obtain from (\ref{eq58})
that $\tau(p,\Theta)=0$ or $p\circ \Theta=p* \Theta.$ Thus, kinetic energy balance (\ref{eq44}) becomes 
\be  \partial_t(\frac{1}{2}\rho v^2)+\grad\bdot[(p+\frac{1}{2}\rho v^2)\bv]=p*  \Theta \lb{eq92} \ee
and internal energy balance (\ref{eq57b}) becomes 
\be  \partial_t u+\grad\bdot(u\bv)=-p* \Theta. \lb{eq93} \ee
In other words, kinetic energy and internal energy balances hold without anomalies for 
flows more regular than allowed by (\ref{eq89-1})-(\ref{eq89-3}). It is interesting that the proof of a complete 
Onsager singularity theorem for energy dissipation anomalies in compressible turbulence seems 
to require an essential consideration of entropy.   
 

\textcolor{black}{
We again close the section by considering steady-state 
compressible turbulence which is forced mechanically and also cooled, e.g. by radiation,
governed by equations (\ref{eq64})-(\ref{eq66}), this time for entropy balances. 
It is easy to see that the equation (\ref{eq57}) for intrinsic resolved internal energy is modified 
by the addition of two terms:  
\be \partial_t\overline{u}^* =\cdots + \bar{\rho}\widetilde{\tau}(\bv; {\bf a}_{ext}) -\overline{Q}_{cool}. \lb{eq93a} \ee 
The first term $\bar{\rho}\widetilde{\tau}(\bv; {\bf a}_{ext})$ is negligible when
$\ell\ll L,$ where $L$ is the length-scale of the smooth acceleration field ${\bf a}_{ext}.$
However, the second term $-\overline{Q}_{cool}$ has a non-zero limit as $\ell\to 0$ and 
contributes to the balance (\ref{eq89b}) of intrinsic resolved entropy an additional 
term $-\ubeta\overline{Q}_{cool}$ on the right, which represents the removal of entropy at 
large scales by cooling. Likewise, the fine-grained entropy balance (\ref{eq19}) gains an additional
term $-\beta Q_{cool}$ on the right. A statistically stationary and homogeneous steady-state 
must thus satisfy 
\be \langle \Sigma^{diss}\rangle = \langle \beta Q^{cool} \rangle \lb{eq93b} \ee 
from the fine-grained balance and 
\begin{eqnarray} 
\langle \Sigma^{inert*}_\ell\rangle &=& \langle \ubeta_\ell \overline{Q}^{cool}_\ell \rangle 
+ \langle\ubeta_\ell \bar{\rho}_\ell\widetilde{\tau}_\ell(\bv; {\bf a}_{ext})\rangle \cr
  & \approx &   \langle \beta Q^{cool} \rangle, \quad \ell\ll L \lb{eq93c} 
\end{eqnarray}  
from the coarse-grained/inertial-range balance \footnote{Because of the assumed smoothness
of $Q_{cool},$ the product $\beta Q_{cool}$ is meaningful in the ordinary sense in the ideal limit 
and no special care is required for its definition.}. The physical picture is that the large-scale negentropy
introduced by cooling cascades through an inertial-range down to small-scales where it is cancelled 
by microscopic entropy production. Alternatively, we may write (\ref{eq93c}) as 
\begin{eqnarray} 
\langle \Sigma^{flux*}_\ell\rangle &=& \langle I^{flux}_\ell\rangle+\langle \ubeta_\ell \overline{Q}^{cool}_\ell \rangle 
+ \langle\ubeta_\ell \bar{\rho}_\ell\widetilde{\tau}_\ell(\bv; {\bf a}_{ext})\rangle \cr
  & \approx &  \langle I^{flux}\rangle+ \langle \beta Q^{cool} \rangle, \quad \ell\ll L\lb{eq93d} 
\end{eqnarray}  
so that the flux of intrinsic entropy through the initial-range includes also negentropy input from
anomalous pressure-work, as well as from large-scale cooling. For an ideal gas equation of state
$I^{flux}_\ell=0,$ of course, and $\langle \Sigma^{flux*}_\ell\rangle\approx  \langle \Sigma^{diss}\rangle$ 
for $\ell\ll L.$} 


\textcolor{black}{
In addition to these exact relations, some approximate balances are suggested by our results. 
For the purpose of a qualitative picture, let us 
assume an approximately constant temperature $T.$
This gives (on average) the fine-grained (dissipation-range) entropy balance 
\be  \langle\Sigma_{therm}\rangle+\frac{\langle Q\rangle}{T}=\frac{\langle Q_{cool}\rangle}{T} \lb{eq94} \ee
and the coarse-grained (inertial-range) entropy balance 
\be  \langle\Sigma_{flux}\rangle + \frac{\langle Q\rangle}{T} = \frac{\langle Q_{cool}\rangle}{T} + \langle I_{mech}\rangle. \lb{eq95} \ee
One also has the mean energy balance $\langle Q_{cool}\rangle =\langle Q\rangle +\langle Q_{trans}\rangle$ 
with $Q_{trans}=-p*\Theta$. Note that the fine-grained pressure-work $Q_{trans}$ is an incoherent transfer of energy,
which might be better denoted in this context as $Q_{incoh}.$  The first equation (\ref{eq94}) thus gives
\be  \langle \Sigma_{therm}\rangle =\frac{\langle Q_{incoh}\rangle}{T}. \lb{eq96} \ee
Decomposing $\langle I_{mech}\rangle = (\langle Q_{coh}\rangle-\langle Q_{incoh}\rangle)/T$ with the definition 
$\langle Q_{coh}\rangle=-\lim_{\ell\rightarrow 0}\langle \up \ \bar{\Theta}\rangle$ of coherent work input,   
substituting into the second equation (\ref{eq95}), and using $\langle Q_{cool}\rangle=\langle Q\rangle +\langle Q_{incoh}\rangle$ 
again yields
\be  \langle \Sigma_{flux}\rangle = \frac{\langle Q_{coh}\rangle}{T}. \lb{eq97} \ee
The above relations implicitly assume that $\langle \beta*  Q_{visc}\rangle\simeq \langle \beta\circ Q_{visc}\rangle
\simeq \langle Q\rangle/T.$ 
This heuristic argument suggests that the coherent input of negentropy at large scales by anomalous pressure work 
will be cascaded downscale, while the positive entropy input due to fine-grained (incoherent) transfer from mechanical to internal 
energy will go into entropy production due to thermal conduction. Needless to say, this is a non-rigorous 
mean-field argument ignoring temperature fluctuations and must be subject to empirical tests in order to determine its 
range of validity. It clearly requires a sufficiently large Mach number, since $Q_{coh}$, $Q_{incoh}$ both vanish
for incompressible flow.  However, like our earlier arguments, it supports the conclusion that compressible turbulence 
will generally involve a forward cascade of negentropy, or, equivalently, an inverse cascade of the traditional entropy.}

\section{Relations to Other Approaches}\label{sec:other}

We now briefly discuss the relation of our analysis with other approaches to compressible fluid turbulence
that have been proposed. 

\subsection{Barotropic Models}\label{sec:barotropic}

Barotropicity is a common assumption that is employed to simplify the description of compressible turbulence,
with pressure $p=p(\rho)$ taken to be a function of mass density only. Polytropic models with $p(\rho)=K\rho^\gamma$ 
are a special case. This assumption has been employed in many classical theoretical papers 
\cite{chandrasekhar1951fluctuations,kraichnan1955statistical}
and also more recent theoretical works 
\cite{erlebacher1990analysis,galtier2011exact,banerjee2014kolmogorov}. 
Barotropic models are often also employed for numerical simulations, especially in astrophysical applications 
\cite{kritsuk2007statistics,kritsuk2013energy}.  No explicit equation for internal energy is solved in such models, and instead 
the internal energy per mass is obtained from 
\be  e(\rho) = \int p(\rho) \frac{d\rho}{\rho^2} \lb{eq98} \ee
so that $D_t\rho = -\rho (\grad\bdot\bv)$ implies
$ \rho D_t e = - p(\grad\bdot \bv)$
and then $u=\rho e$ is recovered satisfying
\be  \partial_t u + \grad\bdot(u\bv)= -p(\grad\bdot\bv). \lb{eq99} \ee
The first law of thermodynamics in the form
\be  de=T ds_m+ \frac{p}{\rho^2}d\rho, \lb{eq100} \ee
with $s_m=s/\rho$ the entropy per mass, shows that  $u$ obtained as above can consistently 
represent internal energy only when either $T=0$ 
(so that $s=0$ by the Nernst law) or else entropy per particle $s_n=s/n$ is constant. Barotropicity and
(\ref{eq98}) are thus equivalent to isentropy.  It follows that the barotropic approximation rules 
out {\it a priori} the entropy cascade central to the present theory.

The fundamental problem is that the assumption of isentropy is in conflict with viscous dissipative
dynamics \footnote{This problem was acknowledged in some of the early works assuming 
isentropy. For example, Chandrasekhar \cite{chandrasekhar1951fluctuations}
admitted that ``It is arguable that the assumption 
of the adiabatic relation ... is not compatible with the retention of the term in viscosity...''}.  
Whereas a smooth ideal Euler solution can be consistently taken to be isentropic, viscous barotropic models 
are physically inconsistent approximations to compressible Navier-Stokes, because they are in conflict 
with conservation of total energy for closed systems! The kinetic energy which is lost by viscous dissipation
should reappear as internal energy of the fluid, but the equation (\ref{eq99}) for the internal energy in barotropic 
models contains no viscous heating. Thus viscous barotropic models make the fallacious prediction that the 
total energy of the isolated fluid is non-conserved.  While fundamentally defective as a model 
of compressible Navier-Stokes, the barotropic approximation is possibly adequate as a crude model 
for astrophysical fluids where \textcolor{black}{kinetic energy lost by cascade is not acquired by the internal energy 
of the considered fluid, e.g. weakly collisional plasmas with negligible viscosities 
\cite{schekochihin2009astrophysical}.}
In addition to the above theoretical problems, the isentropic approximation 
is also observed empirically to be not well satisfied pointwise in subsonic and transonic turbulence for an ideal gas 
\cite{donzis2013fluctuations,ni2015effects}. Substantial entropy fluctuations $s_m=c_V\log(p/K\rho^\gamma)$ 
of the ideal gas are found, which are completely 
neglected by barotropic models and that are very physically significant (as discussed above and more below). 

The polytropic model $p=K \rho $ with adiabatic index $\gamma=1$ requires special discussion, because 
it is often interpreted by means of the ideal gas law $p=nk_B T$ as an isothermal ideal gas, rather
than as an isentropic fluid. With this interpretation the integral $\int p(\rho) \frac{d\rho}{\rho^2} $ 
does not yield the internal energy per mass $e,$ but instead the chemical potential per mass $\mu_m=\mu/m.$
Indeed, the Gibbs-Duhem relation $dp=sdT+\rho d\mu_m$ with $dT=0$ and $d(p/\rho)=0$ 
immediately yields $d\mu_m=pd\rho/\rho^2$. Consistently, substituting $p=(k_BT/m)\rho$ gives 
\be  \int p(\rho) \frac{d\rho}{\rho^2} =(k_BT/m) \log(\rho/\rho_0), \lb{eq101} \ee 
which agrees with the chemical potential per mass of an ideal gas up to an additive constant depending
only upon temperature. To obtain the internal energy consistently within this isothermal interpretation 
of the model one must rather than (\ref{eq98}) instead use 
\be  u= \alpha n k_B T = c_V \rho T, \quad \alpha=1/(\gamma-1) \lb{eq102} \ee
which implies that $u$ is proportional to $\rho,$ just like $p,$ and satisfies the equation
\be  \partial_t u + \grad\bdot(u\bv)=0. \lb{eq103} \ee
Hence, a viscous isothermal gas model is also in conflict 
with conservation of total fluid energy, because kinetic energy lost by viscous dissipation is not transferred
into internal energy. In addition, strict isothermality is not a dynamically consistent assumption for compressible 
Navier-Stokes. The temperature equation can be easily checked for a general equation of state to have the 
form  
\be  \rho c_V D_t T = - T\left(\frac{\partial p}{\partial T}\right)_n \Theta
+ \grad\bdot(\kappa\grad T)+2\eta S^2 +\zeta \Theta^2, \lb{eq104} \ee
where $c_V(n,T)$ and $p(n,T)$ are taken to be functions of the two independent thermodynamic variables $n,$ $T$. 
For an ideal gas this simplies further to 
\be  \rho c_V D_t T = - p \Theta
+ \grad\bdot(\kappa\grad T)+2\eta S^2 +\zeta \Theta^2, \lb{eq105} \ee
with $c_V=\alpha k_B/m$ constant. Clearly, pressure work and viscous heating will produce 
thermal inhomogeneities if none were present initially.
At most it can be true in the presence of strong cooling, which adds a term $-Q_{cool}$ to the right side of (\ref{eq104})
and (\ref{eq105}), that $T'/T_0\ll \rho'/\rho_0,$ i.e. that 
temperature fluctuations are much smaller than density fluctuations. However, as discussed in the previous 
section, strong cooling is also a strong source of negentropy and the standard isothermal models 
provide no representation either of entropy production or of nonlinear dynamics of negentropy cascade 
to offset this. It thus highly questionable whether ``isothermal'' models produce a consistent and accurate 
representation of a strongly cooled compressible fluid with small temperature fluctuations.  

\subsection{Point-Splitting Regularization}\label{sec:splitting} 

As discussed in the Introduction, Onsager derived his result on dissipative anomalies for incompressible 
Euler equations in his original work using a point-splitting regularization 
\cite{onsager1949statistical,eyink2006onsager}, very closely related to the methods used by Kolmogorov 
\cite{kolmogorov1941dissipation}
to derive his statistical $4/5$th-law for incompressible turbulence. Onsager's argument was later made completely rigorous by 
Duchon-Robert \cite{duchon2000inertial} and a formal analogy to point-splitting derivations of anomalies in quantum
gauge theories was pointed out by Polyakov \cite{polyakov1992conformal,polyakov1993theory}.  It is therefore natural to consider 
whether the point-splitting approach can be applied as well to compressible fluids. Two 
different groups of researchers have already attempted to obtain statistical relations analogous 
to the ``4/5th-law'' for compressible turbulence by distinct variations of point-splitting methods: 
Galtier \& Banerjee \cite{galtier2011exact,banerjee2014kolmogorov} and Falkovich-Fouxon-Oz \cite{falkovich2010new}.  
We briefly discuss the works of both of these groups, in the light of our own results. 
 
\vspace{5pt}  
 
{\it Galtier-Banerjee Relations}: These authors have attempted to derive ``4/5th-laws'' for compressible 
turbulence within isothermal \cite{galtier2011exact} and polytropic \cite{banerjee2014kolmogorov} fluid models, 
by a point-splitting of the total fluid energy (sum of kinetic and internal energy). It must  be pointed out firstly 
that the quantity called ``internal energy" for an isothermal fluid
in \cite{galtier2011exact}, $e=\int p(\rho) d\rho/\rho^2,$ is in fact the chemical potential per mass.
See previous section. However, much more seriously, we have demonstrated in section \ref{sec:theory}
that there is no turbulent cascade/dissipative anomaly of total energy in a compressible Navier-Stokes fluid!  
As observed also in the previous section, the viscous isothermal/barotropic models studied 
in \cite{galtier2011exact,banerjee2014kolmogorov} are inconsistent with conservation of fluid energy. 
For these mathematical models, \cite{galtier2011exact,banerjee2014kolmogorov} derive a relation interpreted 
as describing a flux $\varepsilon$ of total energy (kinetic $+$ internal) through the inertial-range, which is finally 
dissipated by viscosity. However, in a Navier-Stokes fluid there is no viscous dissipation  of total energy, 
which is a strictly conserved quantity, and the input $\varepsilon$ of total energy from external sources 
(stirring, cooling, etc.) must identically vanish in a long-time steady state. The results of Galtier-Banerjee 
therefore have no validity for compressible turbulence in a Navier-Stokes fluid. 

The possibility remains that the viscous isothermal/barotropic systems studied in \cite{galtier2011exact,banerjee2014kolmogorov}
may be acceptable as very crude models of weakly collisional astrophysical fluids, in which the 
viscosity is a proxy for some other physical mechanism which transforms the cascaded kinetic energy
at small scales not into internal energy of the fluid in question but into some other form (e.g. internal energy 
of another fluid component, electromagnetic radiation, non-thermal particle acceleration, etc.) 
Note that  the inertial-range analysis of the 
present paper and also of \cite{aluie2011compressible,aluie2013scale} applies to the viscous barotropic model and also to the viscous 
``isothermal'' model, if the later is interpreted as an isentropic ideal gas with adiabatic index $\gamma=1.$ The only 
change to our analysis of the inertial-range balance of $u$ in section \ref{sec:internal} is that terms $Q_{visc}$ 
must be set there to zero. To justify a fluid approximation for internal energy there must be some additional
physical mechanism, not explicitly represented in eq.(\ref{eq99}), which regularizes its solution. For example, 
this could be the tiny thermal conductivity of electrons due to rare collisions with ions
\cite{narayan2001thermal,lazarian2006enhancement}. There may also be very weak collisional heating enhanced 
by kinetic mechanisms \cite{schekochihin2008gyrokinetic, schekochihin2009astrophysical}.  Our analysis leads, 
however, to a very different picture than that of Galtier \& Banerjee \cite{galtier2011exact,banerjee2014kolmogorov} 
for their own models, \textcolor{black}{where we predict no cascade of internal energy}.
Any ``cascade of total energy'' is only via kinetic energy cascade in our analysis.  

It is also interesting to ask whether the results of the present paper on a kinetic energy anomaly for Euler equations 
might be alternatively derived by the Galtier-Banerjee point-splitting. The answer is no. The reason is 
that the point-splitting employed by Galtier-Banerjee is not a proper regularization of the kinetic energy equation 
and does not remove divergences in the infinite Reynolds-number limit. To see this, we note that the key identity 
in \cite{galtier2011exact,banerjee2014kolmogorov} for the point-split kinetic energy evolved under isentropic Euler dynamics is 
\begin{eqnarray}
&& \partial_t(\bj\bdot\bv'+\bj'\bdot\bv) \equiv \grad_\br\bdot \left[(\delta \bj\bdot\delta\bv)\delta\bv\right] 
+h_m'(\grad\bdot\bj)+ h_m(\grad\bdot\bj')\cr
&& \qquad +(\bj\bdot\bv'-\bj\bdot\bv+p)\grad\bdot\bv'+(\bj'\bdot\bv-\bj'\bdot\bv'+p')\grad\bdot\bv.  \cr
&& \lb{eq106} \end{eqnarray}
Here quantities marked with a prime ``$\prime$'' are evaluated at a space point $\bx+\br,$ while unmarked 
quantities are evaluated at point $\bx,$ and $\delta f$ is difference $f(\bx+\br)-f(\bx).$  The notation 
``$\equiv$'' indicates equality up to overall space-gradient terms $\grad_\bx\bdot(...)$ which represent 
space-transport of kinetic energy. Finally, $h_m=e+p/\rho$ is the enthalpy per mass, which satisfies 
$dh_m=dp/\rho$ for isentropic flow. In the original work of Onsager \cite{onsager1949statistical,eyink2006onsager}
and Duchon-Robert \cite{duchon2000inertial}, a coarse-graining operation was applied to the separation-vector $\br$  
to obtain a fully regulated expression. However, if the same approach is applied to the above identity,
one gets terms that are ill-defined in the infinite-Reynolds number limit. For example, the last term 
on the right gives the contribution $(\bbj\bdot\bv-\overline{\mbox{\boldmath{$\j$}}\bdot\bv}+\overline{p})\grad\bdot\bv$
which involves a non-smooth function $\bv$ multiplied with a distribution $\grad\bdot\bv.$ Such terms are 
ill-defined at infinite Reynolds-number. Instead the coarse-graining approach of \cite{aluie2013scale} and the 
present paper yields fully regularized expressions, as in our eq.(\ref{eq41}).  Notice that the terms which cause 
trouble for point-splitting as a regularizer are absent in the incompressible case, because $\grad\bdot\bv=0$

Just to be clear, we are not claiming that there is a mathematical mistake of a trivial sort in the 
analyses of Galtier \& Banerjee \cite{galtier2011exact,banerjee2014kolmogorov}. All of their calculations are meaningful and correct at  
finite Reynolds numbers. In fact, their mathematical relations have been checked to be true in numerical simulations
of supersonic ``isothermal'' turbulence \cite{kritsuk2013energy}. What we are claiming is that there are 
unphysical assumptions underlying the mathematical models employed by Galtier \& Banerjee 
\cite{galtier2011exact,banerjee2014kolmogorov} 
and erroneous physical interpretations of the mathematical results.  \textcolor{black}{
Their failure to regularize UV divergences associated to dissipative anomalies prevents them 
from drawing any conclusions on the infinite $Re$ limit.}

\vspace{5pt}

{\it Falkovich-Fouxon-Oz Relation}: In the paper of Falkovich et al. \cite{falkovich2010new} another generalization
of the ``4/5th-law'' to compressible turbulence has been obtained for a barotropic fluid. 
This approach has also been applied to relativistic fluid turbulence by Fouxon \& Oz \cite{fouxon2010exact}
as we will discuss in a following paper \cite{eyink2017cascadesII}. 
Consideration of a point-split quantity $\bj\bdot\bj'$ allowed \cite{falkovich2010new}
to derive an exact 
relation for homogeneous, isotropic statistics, which reduces to the standard 4/5th-law 
in the incompressible limit. The quantity which is cascaded to small scales in their picture 
is the input of $(1/2)|\bj|^2$ by external forcing. The exact equation obeyed by this field 
for a smooth solution of compressible Euler equations (without need of any barotropic assumption) is
\be  \partial_t\left(\frac{1}{2}|\bj|^2\right)+\grad\bdot\left(\frac{1}{2}|\bj|^2\bv+ p \bj\right)
= -\frac{1}{2}|\bj|^2(\grad\bdot\bv) + p(\grad\bdot\bj). 
\lb{eq107} \ee
As a matter of fact, it is not hard to show that this balance equation may indeed be anomalous 
in a high-Reynolds-number compressible turbulence and to use a point-splitting regularization 
to derive the anomaly. With the same notations as in eq.(\ref{eq106}), one easily finds 
\begin{eqnarray}
&& \partial_t(\bj\bdot\bj') \equiv \frac{1}{2}\grad_\br\bdot \left[ |\delta \bj|^2\delta\bv\right]  \cr
&& \qquad  -\frac{1}{2}|\bj|^2(\grad\bdot\bv') -\frac{1}{2}|\bj'|^2(\grad\bdot\bv) + p(\grad\bdot\bj') + p'(\grad\bdot\bj). \cr
&& \lb{eq108} \end{eqnarray} 
Unlike the previous case, all terms are fully regularized after coarse-graining in 
the separation-vector $\br$ and one obtains an anomaly term $-A$ appearing on the right  side
of eq.(\ref{eq107}) for infinite Reynolds number, with 
\be  A={\mathcal D}\mbox{-}\lim_{\ell\rightarrow 0} \frac{1}{4\ell}\int d^dr (\grad G)_\ell(\br)\bdot
\delta\bv(\br) |\delta \bj(\br)|^2. \lb{eq109} \ee 
It is also straightforward to derive the anomalous balance 
equation for $(1/2)|\bj|^2$ by using the coarse-graining approach of the present paper, but we leave this as 
an exercise for the reader. Note that in this balance equation one faces the same issue of 
defining products like $\frac{1}{2}|\bj|^2\circ\Theta$ and $\frac{1}{2}|\bj|^2*\Theta,$ similar to  
pressure-work in the energy balances in section \ref{sec:energy} of the present paper. 

As with the previous point-splitting approach, we conclude that the result of Falkovich et al.
\cite{falkovich2010new}
is mathematically correct and, even more, the derivation is valid in the infinite Reynolds-number limit. 
The statistical relation of \cite{falkovich2010new} has also been verified in a numerical simulation 
of ``isothermal'' compressible turbulence \cite{wagner2012flux} (although there are some 
subtle issues in the statistical evaluation of the external input). 
However, we disagree completely with the conclusion that the result of \cite{falkovich2010new}
``...indicates that the interpretation of the Kolmogorov relation for the incompressible 
turbulence in terms of the energy cascade may be misleading'' (Fouxon \& Oz, \cite{fouxon2010exact}). 
Such a conclusion could be justified if that relation were the only possible 
generalization of the 4/5th-law to compressible turbulence. 
However, the analysis of \cite{aluie2011compressible,aluie2013scale} and ours in section \ref{sec:kinetic}
of the present paper fully support the existence of a kinetic energy cascade for compressible 
turbulence and yield the analogue of 4/5th-laws for the kinetic energy flux. A further issue 
with the result of Falkovich et al. \cite{falkovich2010new} is that we see no compelling 
interest in the quantity $(1/2)|\bj|^2$ for compressible fluids. It is neither a conserved quantity 
nor any component of a conserved quantity, and it has no obvious dynamically important 
role in compressible turbulence. Just as in quantum field-theory, it is not hard to find infinitely many anomalous 
balance relations in the ideal limit of turbulence but most of them are not physically relevant and 
have no significant consequences. In our opinion, the deep importance of the 4/5th-law for 
incompressible turbulence arises from its connection to the dissipative anomaly for kinetic 
energy and its implication that fluid singularities of the type $\zeta_q^v\leq q/3$ are required for 
such an anomaly. Our analysis shows that such a connection fully extends to compressible fluid turbulence. 

\vspace{5pt}

As our final comments in this section, we would like to emphasize the general limitations of point-splitting 
in turbulence theory. It is not ruled out by the analysis in this paper that a clever point-splitting 
may someday be found for compressible turbulence which will yield the anomalous kinetic energy balance (\ref{eq60}). 
However, it is very hard to imagine that a point-splitting regularization will ever be found to yield the anomalous 
entropy balance (\ref{eq76}) in this paper. 
The coarse-graining approach that we employ is a more powerful and general method than point-splitting. 
In addition to the Eulerian balances discussed here, coarse-graining can also be employed to obtain Lagrangian 
conservation-law anomalies, such as for fluid circulation in hydrodynamic turbulence 
\cite{eyink2006turbulent} and magnetic flux conservation in MHD turbulence
\cite{eyink2006breakdown,eyink2015turbulent}.
 
\subsection{Decomposition into Linear Wave Modes}\label{sec:linear}

Another common theoretical approach to compressible turbulence, which goes back to work of Kov\'{a}sznay 
\cite{kovasznay1953turbulence} and Chu \& Kov\'{a}sznay \cite{chu1958nonlinear}, 
is to expand compressible Navier-Stokes solutions into linear wave modes, 
based on an assumption of small perturbations around a homogeneous state and weak nonlinearity.  This 
expansion identifies three ideal linear wave modes \footnote{For finite values of $\eta,$ $\zeta,$ $\kappa,$
the modes have complex frequencies with imaginary parts reflecting dissipative decay. For example,
see  \cite{chu1958nonlinear},  equations 6.1-3. Here we focus on the ideal nonlinear behavior 
at very high Reynolds and P\'eclet numbers, as  in most of  \cite{chu1958nonlinear}, section 6.}, 
the ``sound mode'' of frequency $c_sk$ for sound-speed
$c_s$ and wave-number $k,$ and  two zero-frequency modes, the 
``vorticity mode'' and the ``entropy mode.'' An obvious question, which we address here, is how the 
``entropy mode'' of Kov\'{a}sznay is related to our concept of an entropy cascade.  

To briefly review the approach of Kov\'{a}sznay \cite{kovasznay1953turbulence} and Chu \& Kov\'{a}sznay 
\cite{chu1958nonlinear}, we recall that  
that it assumes an ideal gas equation of state, with pressure and entropy per particle given by  
\be p=n k_B T, \qquad s_n=k_B \log(T^\alpha/ C n) \lb{eq110} \ee 
as functions of $n$ and $T,$  with $\alpha=1/(\gamma-1).$ Linearization around a homogeneous state
satisfying $p_0=n_0 k_B T_0$ yields for the fluctuations the linear relations
\be  \frac{p'}{p_0}=\frac{n'}{n_0}+\frac{T'}{T_0}, \qquad \frac{s_n'}{k_B}= -\frac{n'}{n_0}+\alpha \frac{T'}{T_0}. \lb{eq111} \ee
Here we use the prime ``$\prime$'' to denote a putatively small fluctuation value.  
For ideal flow the ``sound mode'' has $s_n'=0$ and the ``entropy mode'' has $p'=0.$
(For non-ideal flow Kov\'{a}sznay  finds instead a small entropy $s_n'$ associated to the ``sound mode'', 
which is proportional to the molecular transport coefficients or dimensionless ``Kundsen number'' $\epsilon$
and which is neglected at zeroth order in $\epsilon$). The zeroth-order dynamics of the fluctuations 
for ideal flow are found to be given by the linear equations 
\be  \partial_t\bomega'=\bzed, \quad \partial_t s_n'=0, \quad \partial_t^2p'-c_s^2 \nabla^2 p'=0\lb{eq112} \ee 
with $\bomega'=\grad\btimes\bv'$ the vorticity fluctuation.
See \cite{chu1958nonlinear}, eq.(6.5). Nonlinearity is recovered in the Kov\'{a}sznay approach 
by expansion to second-order in the nonlinearity, which yields mode-mode coupling terms, such as 
vortex self-stretching (a vorticity-vorticity mode coupling). See Table 1 of \cite{chu1958nonlinear}
for a complete tabulation of all second-order interactions. The only such couplings that contribute to entropy dynamics 
are entropy-vorticity and entropy-sound couplings of the form $-\bv' \bdot \grad s_n'$, which describe advection 
of entropy by velocity fluctuations $\bv'$ due to vorticity and sound modes. Thus to quadratic order 
in nonlinearity, the entropy per particle $s_n'$ appears as a passive scalar and 
entropy per volume $s'=n s_n'$ as a passive density. 

Independent of our work, there are a number of serious problems with the Kov\'{a}sznay modal 
expansion when considered as an {\it a priori} theoretical approach. First and foremost, there 
is no small parameter on which to base such an expansion. Instead, fluctuations of thermodynamic variables 
in compressible flow can be very large relative to mean or r.m.s. values, as seen for example in 
\cite{donzis2013fluctuations}, Fig.~4. This essential strong-coupling nature is, of course, the most well-known 
theoretical difficulty with the analysis of turbulent flow. 
A closely related problem is that solutions of the compressible Navier-Stokes equation cannot 
be consistently expanded into linear wave modes, because there is no superposition principle 
for such nonlinear dynamics. Even for an ideal gas, the thermodynamic relations 
(\ref{eq110}) impose nonlinear constraints between $p,n,T$ or $s_n,n,T,$ which will not be satisfied 
for superpositions of wave modes except in the very crude linear approximation (\ref{eq111}). 
For second-order moments of $p'/p_0$ and modest Mach numbers (0.1-0.6) the 
predictions of the linear approximation (\ref{eq111}) are adequate to about 1\% level (see data in Table 1 
and Figs.2-3 in \cite{donzis2013fluctuations}), but the error grows with increasing Mach number 
and also for higher moments/larger fluctuations. 
It is worth noting that Kov\'{a}sznay himself was not attempting in his original works to 
develop a general theoretical approach for analysis of compressible fluid turbulence, but his goal 
was instead a more modest one of constructing a decomposition to assist in the interpretation of 
experimental measurements. 
Some later researchers have taken this type of modal decomposition much more literally than it was
first intended. 

Our analysis in this paper has shown that the entropy in high Reynolds-number compressible
turbulence is not at all a passive scalar. Entropy is, of course, a nonlinear function of basic thermodynamic
variables, e.g. $s(u,n)$ taken as a function of internal energy density $u$ and particle density $n.$ Its 
dynamics is completely determined by the dynamics of $u$ and $n$ and it is, in that sense, ``passive''. 
However,  the turbulent dynamics of entropy in the ideal limit of vanishing molecular transport is 
not that of a passive scalar. Comparing the weakly nonlinear expansion result 
\be  \partial_t s' + \grad\bdot(s'\bv')=0. \lb{eq113} \ee
with our own eq.(\ref{eq76}), we see that, beyond passive advection, \textcolor{black}{the inertial-range dynamics 
of entropy involves both anomalous input $I_{mech}$ of negentropy from 
pressure-work and nonlinear entropy cascade $\Sigma_{flux}$ (as well as entropy production by viscous heating). 
If one tried to interpret the mechanical input $I_{mech}$ of 
entropy crudely within the Kov\'{a}sznay framework, it would have to be considered a turbulent ``sound-sound" 
coupling which produces negentropy.  It is completely missed by the Kov\'{a}sznay  weakly nonlinear expansion 
which cannot detect such ``anomalous" terms.  Finally, the identification of $s_n'$ as a passive scalar would 
imply that there is a forward cascade of $|s_n'|^2,$ but our analysis instead predicts an inverse cascade 
of the entropy $s$ as a nonlinear function of $u$ and $n.$ Our predictions for entropy are thus fundamentally
different from those obtained by treating the linear ``entropy mode'' as a passive scalar.}   

\section{Empirical Consequences and Evidence}\label{sec:empirical}

\textcolor{black}{
Our analysis yields a great many predictions testable by laboratory experiments and numerical 
simulations, the two most novel being the pressure-dilatation defect $\tau_\ell(p,\Theta)$ contribution
to anomalous kinetic energy dissipation and the anomalous production of negentropy by 
pressure-work $I^{flux}_\ell$ and nonlinear negentropy cascade $\Sigma^{flux*}_\ell.$ These quantities 
are all straightforward to calculate in simulations of compressible turbulence, where inertial-range
contributions such as $\bar{\tau}_\ell(p,\Theta)$ or negentropy flux $\Sigma_\ell^{flux*}$ can be obtained 
by numerical implementation of the spatial filtering.} Laboratory experiments can also measure 
such quantities using techniques such as holographic PIV \cite{tao2002statistical}.  Our analysis yields 
as well testable predictions on scaling exponents through the inequalities (\ref{eq89-1})-(\ref{eq89-3}). 
Here we may note in particular the predictions for ``roughness'' of the internal energy density and mass density fields 
in order for an entropy cascade to exist, with structure-function exponents $\zeta_q^u$ and $\zeta_q^\rho$
required, essentially, to be less than or equal to K41 values for $q\geq 3.$ Not only are these various 
predictions able to be checked in detail in future studies, but also many past works in retrospect provide 
supporting evidence. We next discuss some of this prior work.  

First, the previous numerical studies of the pressure-work \cite{aluie2012conservative,ni2015effects}
provide evidence for a pressure-work defect $\tau(p,\Theta)$, although this was not clearly understood at the time. 
The main object of those studies was the saturation of $\langle \bar{p}_\ell\bar{\Theta}_\ell\rangle$ for
$\ell$ decreasing through the inertial-range. This was demonstrated through study of the pressure-dilatation
cospectrum  
\be  PD(k) = - \sum_{{\bk}':\  ||{\bk}'|-k|<0.5 } \hat{p}(\bk')\hat{\Theta}(-\bk') \lb{eq114} \ee
and of the statistics of the pressure-dilatation residual $p\Theta-\bar{p}\bar{\Theta}.$ In both studies 
\cite{aluie2012conservative,ni2015effects} it was found that the cospectrum exhibited a power-law behavior $PD(k)\sim C k^{-\beta}$ 
in the inertial range, crucially with $\beta>1$ so that the integral over the range $k\in [0,\infty)$ would converge. 
However, in the finite Reynolds number simulations the power-law with exponent $\beta$ persists only 
over a finite range and in the dissipation-range the cospectrum was found to lie {\it above} the inertial-range power-law.
See, \cite{aluie2012conservative}, Fig.~2 and \cite{ni2015effects}, Fig.~23. 
This is the signature to be expected from a positive mean defect $\langle \tau(p,\Theta)\rangle >0.$
Even more relevant are the previous numerical results for the pressure-dilatation residual, since it is 
directly related to the mean defect by 
 \be  \langle \bar{\tau}(p,\Theta) \rangle = \langle p\Theta-\bar{p}\bar{\Theta} \rangle. \lb{eq115} \ee
The two previous studies both found that the residual for $\ell$ near 
the bottom of the inertial-range took on very large positive and negative values associated to shocks 
(small-scale shocklets or large-scale shocks, depending upon the compressibility of the forcing). 
See \cite{aluie2012conservative}, Fig.~4 and \cite{ni2015effects}, Figs.~26,27. The large values nearly cancelled 
in a global space average, leaving only a small positive average $\langle \tau(p,\Theta)\rangle,$ about 
20 times smaller than the asymptotic value $\langle p\circ \Theta\rangle$. Aluie et al. 
\cite{aluie2012conservative} considered this 5\% contribution to be ``negligible''. However,
both the simulations \cite{aluie2012conservative,ni2015effects} were for subsonic and transonic 
turbulence. If $\langle \tau(p,\Theta)\rangle$ arises mainly from shock heating, then it is 
reasonable to expect that this average will make an increasingly large contribution to the kinetic
energy dissipation anomaly for increasing Mach numbers. 

There is also evidence from prior studies for a negentropy cascade. Motivated by incompressible fluid 
turbulence where the temperature is a passive scalar, Ni et al. \cite{ni2015numerical} and Ni \& Chen \cite{ni2015effects}
(see section 6 of both papers) have numerically studied  ``temperature cascade'' in subsonic and 
transonic compressible turbulence of an ideal gas. Using the same coarse-graining approach as the present paper, 
those authors attempted to derive a balance equation for the quantity $G=(1/2)\bar{\rho}\tilde{T}^2$. 
Their result (\cite{ni2015effects}, eqs.(6.3)-(6.8)) contains several errors \footnote{The most serious errors
in the results of \cite{ni2015effects} are in their eqs.(6.6)-(6.8), which correspond to the last three terms on 
the righthand side of our eq.(\ref{eq116}). In particular, the pressure work $\overline{p\Theta}$ in their eq.(6.6)
is factorized as $\overline{p}\overline{\Theta}$ and, most seriously, $\overline{Q}_{visc}$ in their eq.(6.7) is replaced 
with $2\eta|\overline{\bS}|^2+\zeta|\overline{\Theta}|^2.$ This latter quantity vanishes in the ideal limit of 
high Reynolds numbers, whereas the quantity $\overline{Q}_{visc}$ is expected to have a non-zero value} but the 
expression for the subscale flux of $G$ that they obtained is the same as that for the correct equation 
derived from (\ref{eq105}) and given here:
\begin{eqnarray}
&& \partial_t \left(\frac{1}{2}\bar{\rho}\tilde{T}^2\right) + \grad\bdot\left(
\frac{1}{2}\bar{\rho}\tilde{T}^2\tilde{\bv}+\bar{\rho}\tilde{\tau}(T,\bv)\tilde{T}-(\overline{\kappa\grad T})\tilde{T}/c_V\right) \cr
&& \quad = - \Pi^G_\ell + \left(-\overline{p\Theta}+\overline{Q}_{visc}\right)\tilde{T}/c_V
-\grad\bdot(\overline{\kappa\grad T})\bdot\grad\tilde{T}/c_V \cr
&& \lb{eq116} \end{eqnarray}   
with 
\be  \Pi^G_\ell=-\bar{\rho}\grad\tilde{T}\bdot\tilde{\tau}(T,\bv) \lb{eq117} \ee 
the subscale flux of $G$. Studies \cite{ni2015numerical,ni2015effects} have verified numerically that this quantity 
has a positive average $\langle \Pi^G_\ell\rangle>0$ over a range of $\ell,$ indicating a forward cascade 
of the quantity $G$ to small-scales. This is almost direct evidence for a forward negentropy cascade. 

The quantity $G=\frac{1}{2}\bar{\rho}\tilde{T}^2$ was, in fact, first introduced by Obukhov 
\cite{obukhov1949temperature} for incompressible 
fluid turbulence as an approximation to the ``negentropy'' or ``information'' introduced by an ordered temperature 
field, assuming an isobaric ideal gas and small amplitudes of temperature fluctuations
\footnote{Alternatively, Obukhov \cite{obukhov1949temperature} showed that the quantity $G$ is an approximation 
to the available ``free energy'' which can be reversibly extracted from the temperature}.  
\textcolor{black}{The concept of an ``entropy cascade'' was later invoked
by L'vov \cite{lvov1991spectra} for the cascade of $G$ within the Bolgiano-Obukhov picture of convective turbulence
in a Boussinesq fluid. The ideas of Obukhov \cite{obukhov1949temperature}
and L'vov \cite{lvov1991spectra} are the closest analogue 
for an incompressible fluid of the entropy cascade proposed in this work. An important difference  
is that in the theories of  \cite{obukhov1949temperature, lvov1991spectra} the flux $\Pi_\ell^G$ is in statistical 
balance with the ``temperature dissipation'' by thermal conductivity, or $\kappa |\nabla T|^2/c_P,$ without 
any contribution from entropy production due to viscous heating and with no consideration of an 
explicit cooling mechanism. 
For compressible turbulence there is little reason to consider the approximation $G$ rather 
than the correct large-scale entropy $s(\bar{u},\bar{\rho})$ and our balance equations (\ref{eq70})
for $\us$ and (\ref{eq89b}) for $\us^*$ are more theoretically tractable than (\ref{eq116}) for $G$, because entropy is a 
conserved quantity for smooth solutions of compressible Euler equations whereas $G$ is not. 
However, the observation of \cite{ni2015numerical, ni2015effects} that $\langle \Pi^G_\ell\rangle>0$ 
strongly suggests that $\langle \Sigma_\ell^{flux}\rangle>0$ will hold over a similar range of $\ell$
and makes it vital to subject the latter prediction and the balance relations (\ref{eq93c})-(\ref{eq93d}) and 
(\ref{eq96})-(\ref{eq97}) to detailed empirical tests.} Here we note that the spectra of density, temperature, and 
pressure (or, equivalently for an ideal gas, internal energy) in the simulations of \cite{donzis2013fluctuations}  
and \cite{ni2015numerical} are consistent with the roughness expected for negentropy cascade. In particular, for 
transonic Mach numbers ($Ma\approx 0.6$) all three thermodynamic variables have Fourier spectra close to 
$k^{-5/3},$ scaling with the K41 exponent  \footnote{Our rigorous inequalities for structure-function 
exponents are all valid for orders $q\geq 3,$ where the values must be sub-Kolmogorov. As usual, the 
exponents for $q\leq 3$ are then expected to have super-Kolmogorov values, because of the 
concavity of the scaling exponents in the variable $q$ \cite{uriel1995turbulence}}. 


The negentropy cascade proposed here, if correct, must occur for compressible turbulent flows in Nature,
with one of the most significant examples being turbulence in the interstellar medium (ISM). The electron 
density of the ISM exhibits a spectrum close to the Kolmogorov $k^{-5/3}$ over a 13-decade 
range, as inferred from electron scintillation measurements over $10^5$-$10^{10}$ km scales and 
from other observations over $10^2$-$10^{15}$ km 
\cite{armstrong1981density,armstrong1995electron,lazio2004microarcsecond,chepurnov2010extending}. 
The spectacular extent of this scaling range has led the density spectrum 
to be dubbed the ``Big Power Law in the Sky". Because the ion mean free path in the ISM is $\sim 10^7$ km, 
a fluid approximation is expected to be valid over the majority of this range. Magnetic fields also play a 
significant role in the dynamics of the ISM, so that the dynamics of the ISM at length scales above
$\sim 10^7$ km is expected to be that of a compressible magnetohydrodynamic (MHD) fluid with a Mach 
number of order unity.  Our work suggests an identification of the 
``Big Power Law in the Sky" as resulting from a nonlinear inverse cascade of entropy 
(or forward negentropy cascade). Note that all of our results in this paper extend straightforwardly to compressible 
MHD (for which see Landau \& Lifschitz \cite{landau2013electrodynamics}, Chapter VIII, \S 65-66). The only difference is that
now there is a cascade of total mechanical energy (kinetic + magnetic) 
and the corresponding energy dissipation anomaly now contains a contribution from resistive heating 
\be  \bar{Q}=\lim_{\eta,\zeta,\kappa,\gamma\rightarrow 0} \overline{2\eta|\bS|^2+\zeta \Theta^2 + \gamma J^2/4\pi} \lb{eq118} \ee 
where $\gamma=c^2/4\pi\sigma$ is the magnetic diffusivity and $\bJ=\grad\btimes\bB$. 
\textcolor{black}{In particular, our balance eq.(\ref{eq70}) for $\us$ remains valid for compressible MHD 
with the above change to $Q_{diss},$ and the balance eq.(\ref{eq89b}) for $\us^*$ now has contributions 
to $Q^{flux}_\ell$ from the Lorentz force \cite{aluie2010scale}. There is thus by our arguments a forward negentropy cascade
in compressible MHD turbulence. 
We theorize that this nonlinear negentropy cascade is the origin of the plentiful density fluctuations in the large-scales
of the ISM where compressible MHD is valid.}  

Most current theories of the electron density spectrum of the ISM, by contrast, have been developed 
within Kov\'{a}sznay-type linear wave-mode picture for compressible MHD, where the 
basic waves are now the ``shear Alfv\'en mode'', the ``slow magnetosonic mode'', the ``fast 
magnetosonic mode'', and the ``entropy mode'' (e.g. see \cite{kulsrud2005plasma}, Ch.5). In particular, 
one popular theory of the power-law spectrum is that it results from a forward cascade of the ``entropy mode'' 
as a passive scalar \cite{higdon1984density,lithwick2001compressible}. However, the large scales of the ISM are 
believed to be nearly isothermal above a cooling scale $L_{cool}\sim 10^{12}$ km, because of efficient 
radiative cooling (e.g. by electron impact excitation of metal line transitions). Referring to eq.(\ref{eq111}), one sees 
that there can then be no ``entropy mode'' with $p'/p_0=0$ because $T'/T_0=0$ and so cannot cancel 
the density fluctuation $n'/n_0$ (or, more accurately, the entropy mode is 
extremely damped, because $T'/T_0\ll 1$). In that case, the only remaining mode to carry density 
fluctuations is the isothermal sound mode (slow magnetosonic) with $p'/p_0=n'/n_0=-s_n'/k_B$ (which,
in contrast to the adiabatic sound mode, carries entropy fluctuations due to density changes) \footnote{
To make this argument correctly requires a discussion of the linear wave modes of compressible MHD.
This problem is carefully treated by  \cite{lithwick2001compressible}, Appendix A, including also the important 
effects of cooling. As discussed by those authors, the MHD entropy mode with cooling has at sufficiently small scales
zero perturbation not of thermal pressure, but of total pressure (thermal + magnetic). Nevertheless, their 
analysis shows that the above argument based upon the ``entropy mode''
of a hydrodynamic (non-magnetized) fluid carries over to MHD with only minor changes. In fact, 
the hydrodynamic treatment of the entropy mode is exactly valid at sufficiently small-scales where 
slow magnetosonic waves can create pressure balance of the entropy mode, so that $p'=0.$ At larger 
scales the entropy modes carries a non-vanishing fluctuation $p'\neq 0$ of thermal pressure, which is 
balanced by a magnetic pressure fluctuation. In either case, the entropy mode is rapidly damped on a cooling 
time-scale $t_{cool},$ so that the amplitude of the entropy mode can be argued to be small above a cooling scale 
$L_{cool}$. See \cite{lithwick2001compressible}, Appendix A, for all details.}.   
This linear analysis of the fluctuations leads to the so-called {\it cooling catastrophe}, which is concisely 
summarized in this quote: 

\begin{quotation}
\noindent 
``However, the entropy mode is rapidly damped in isothermal turbulence. As a consequence, small-scale density 
fluctuations may be significantly suppressed. There are two possible solutions to this ``cooling catastrophe'': 
either (1) the outer scale is extremely small, small enough that the turbulence at the outer scale is nearly adiabatic ; 
or (2) there are significant density fluctuations associated with the slow mode. However, in the latter case, 
the mean magnetic field must be amplified almost to equipartition with the gas pressure, so that $\beta\sim 1$. 
Either of these two solutions would place stringent constraints on the nature of the turbulence that is responsible 
for observed density fluctuations.'' --- Lithwick \& Goldreich \cite{lithwick2001compressible}  
\end{quotation}

\noindent Within a Kov\'{a}sznay-type modal picture, the slow magnetosonic mode seems the most plausible 
source of the observed density fluctuations. However, in our nonlinear theory,  there is no ``cooling catastrophe" 
in the first place! Large-scale cooling adds excess negentropy (deficiency of entropy) 
\textcolor{black}{that feeds the cascade of negentropy to small scales. See eq.(\ref{eq93c}). This necessitates 
``rough'' density and temperature fields with Kolmogorov-type spectra.}  
In our view, the ``cooling castrophe'' is an artifact of attempting to describe nonlinear compressible MHD 
turbulence in terms of linear wave modes. There is no sound theoretical basis for such a decomposition
and, unsurprisingly, the Kov\'{a}sznay mode-mode interactions lead to empirically wrong predictions for the problem 
\footnote{One example of a failure of the Kov\'{a}sznay mode-coupling theory which has already been considered 
is its prediction that the entropy per particle $s_n'$ is a passive scalar. As another example, consider the estimate 
by \cite{lithwick2001compressible}, section 5.2,  for the spectrum of density at $Ma=1$ in a regime (``high-$\beta$'') where 
magnetic pressure is small relative to thermal pressure. Invoking Kov\'{a}sznay's quadratic ``sound-sound'' coupling
$\partial^2({v'}_i {v'}_j)/\partial x_i\partial x_j,$ they predicted that density perturbations due to isothermal sound waves 
will lead to a $k^{-7/3}$ density spectrum. Instead, \cite{kowal2007density} in a simulation of ``isothermal'' compressible 
MHD at $Ma\sim 1$ have observed for large $\beta$ a $k^{-5/3}$ density spectrum, and \cite{kim2005density} for 
``isothermal'' hydrodynamic ($\beta=\infty$) turbulence at $Ma\sim 1$ found also a $k^{-5/3}$ density spectrum,
contradicting mode-coupling predictions.}.  

A complete presentation of this theory of the ISM electron density spectrum will be given elsewhere, as it 
requires more specialized discussion of MHD turbulence and even plasma kinetics. A very interesting question 
is how our theorized negentropy cascade proceeds to smaller length-scales below the ion mean free path 
where a fluid approximation breaks down. As discussed earlier, the observed $k^{-5/3}$ density spectrum 
in the ISM extends many decades below the ion mean-free path length. The key concept of plasma kinetic turbulence 
is the cascade of negative kinetic entropy or ``free-energy'' (electromagnetic energy minus kinetic entropy) to small
scales of length and velocity in the 1-particle phase space \cite{schekochihin2008gyrokinetic,schekochihin2009astrophysical}. 
The natural conjecture is that the negentropy cascade of compressible MHD turbulence merges with the kinetic cascade at scales 
below the mean-free path, but details remain to be understood. 

%

\section{Discussion}

The theory developed in this paper is based upon the hypothesis that compressible fluid 
turbulence should exhibit dissipative anomalies of energy and entropy, similar to those 
observed for incompressible fluids. From this hypothesis alone, we have argued that 
the high Reynolds- and P\'eclet-number limit should be governed by distributional or ``coarse-grained" 
solutions of the compressible Euler equations. The argument closely follows that of Onsager
\cite{onsager1949statistical,eyink2006onsager} for incompressible fluids, which we have explained as a 
non-perturbative application of the principle of renormalization-group invariance. The theory makes a 
great many predictions that are testable by experiment and simulations, in particular: (1) anomalous 
dissipation of kinetic energy by local energy cascade and by pressure-work defect; 
(2) anomalous input of negentropy into the inertial-range of compressible fluid turbulence  
by pressure-work, in addition to any external input by large-scale cooling mechanisms; (3) negentropy 
cascade to small-scales through a flux of intrinsic inertial-range entropy; and (4) fluid singularities 
required to sustain cascades of energy and entropy, so that at least one of (\ref{eq89-1})-(\ref{eq89-3}),
must hold. 

It should be stressed that even for incompressible fluids, many difficult mathematical questions remain 
open concerning Onsager's theory of ``ideal turbulence'' described by dissipative Euler solutions and its 
main support arises from successful agreement with a broad array of numerical simulations and 
laboratory experiments.  The convex integration theory \cite{delellis2012h,delellis2013continuous} has revealed 
that the Cauchy problem for incompressible Euler equations has non-unique dissipative solutions with fixed initial-data, 
suggesting that the infinite-Reynolds turbulent solutions are essentially unpredictable. So far, no dissipative
Euler solutions of the type conjectured by Onsager have been mathematically derived from incompressible 
Navier-Stokes solutions by the physical limit of vanishing viscosity/infinite Reynolds-number. Work on toy 
``shell models'' suggests that this limit will be very subtle and that the limiting Euler solutions will be non-unique 
and stochastic \cite{mailybaev2015stochastic,mailybaev2016spontaneous,mailybaev2016spontaneously}. 
Further surprises and new insights are doubtless in store. However, Onsager's theory for incompressible 
fluid turbulence has much more empirical support than many other highly-regarded physical theories, 
e.g. Einstein's theory of general relativity. 

For the compressible theory that we have developed here, further work is also clearly required on a 
few key issues.  One of these is the Mach-number dependence of the various physical 
quantities in our theory.  All of our derivations are formally independent of Mach number, but there 
is an implicit Mach-number dependence through the assumption that mass-density $\rho$ remains a 
bounded function in the ideal limit. Instead, there is empirical evidence from numerical simulations 
that for a sufficiently high Mach number the density is not even square-integrable and its 
ideal limit may exist only as a singular measure \cite{kim2005density}. At small Mach numbers the 
anomalous negentropy input by pressure-work must tend to zero. If there is an external heating/cooling 
source to introduce internal energy (or temperature) inhomogeneities at large-scales, then the low Mach-number 
limit of our negentropy cascade must recover that long ago predicted by Obukhov 
\cite{obukhov1949temperature} for 
incompressible fluids. However, if there is no such external source, then our predicted negentropy cascade
presumably disappears for small Mach numbers, but the details are unclear. This is an urgent matter for
evaluating the theory, since much empirical data exists for subsonic and transonic flows. 

A second very important open issue 
has to do with the extension of our theory to kinetic regimes. Our theorem on turbulent entropy 
dissipation anomalies and entropy cascade applies to any distributional solution of compressible 
Euler equations, including those resulting from a kinetic equation. However, it is very unclear 
how our fluid negentropy cascade will merge into a kinetic description at scales much smaller 
than the mean-free-path of the fluid. This is a particularly important issue for plasma kinetics in   
astrophysics \cite{schekochihin2008gyrokinetic,schekochihin2009astrophysical}, because the large 
mean-free-paths frequently encountered in astrophysical plasmas imply that long ranges of scales are 
described by Vlasov-Landau kinetic theory rather than a fluid description. 

One strength of our theory is that it extends readily to relativistic fluid turbulence. This is 
the subject of our following paper \cite{eyink2017cascadesII}.

\begin{acknowledgments}
We thank Ethan Vishniac for very helpful discussions of the physics of the interstellar medium 
and Hussein Aluie for sharing with us his unpublished work. 
\end{acknowledgments}

\appendix

\section{Analytical Shock Solution}\label{Shock}

\subsection{Model and Shock Solution}\label{sec:shocksol}

We shall consider a family of shock solutions derived by Becker \cite{becker1922stosswelle} and 
Johnson \cite{johnson2013analytical,johnson2014closed} for the 1D compressible Navier-Stokes system, obtained 
by reduction of the 3D equations to a single space dimension, with $x$ the distance perpendicular 
to the planar shock and with $v=v_x$ the corresponding velocity component: 
\be  \partial_t\rho+\partial_x(\rho v)=0, \lb{eqA1} \ee
\be  \partial_t(\rho v) +\partial_x(\rho v^2+p-\eta \partial_xv)=0, \lb{eqA2} \ee
\be  \partial_t(\frac{1}{2}\rho v^2 + u) + \partial_x ( \rho v(\frac{1}{2}v^2 +h_m) -\eta \partial_x v-\kappa \partial_x T)=0. \lb{eqA3} \ee
Here $\eta=(4/3)\eta_{3D}+\zeta_{3D}$ for the 3D shear viscosity $\eta_{3D}$ and bulk viscosity $\zeta_{3D}$ \footnote{Our notation 
differs from that of Johnson \cite{johnson2013analytical,johnson2014closed}, with our $\eta$ equal to Johnson's $(4/3)\mu$}.  
An ideal-gas equation of state is assumed, with 
\be  p= (\gamma-1)u, \quad u= c_V \rho T, \quad h_m = c_P T \lb{eqA4} \ee
for any adiabatic index $\gamma=c_P/c_V>1$. The solutions obtained are for the stationary equations,
with all time-derivatives set to zero, and they reduce in the ideal limit ($\eta,$ $\kappa\rightarrow 0$) to stationary shocks 
with discontinuous, step-function solution fields:   
\be  f(x) = \left\{\begin{array}{ll}
                      f_0 & x<0 \cr
                      f_1 & x>0 
                      \end{array} \right. = f_0 + (\Delta f) \theta(x) \lb{eqA5} \ee
Here pre-shock values are labeled by 0 and post-shock values by 1, 
$\Delta f=f_1-f_0,$ and $\theta(x)$ is the Heaviside step function. We also denote 
$f_{av}=\frac{1}{2}(f_0+f_1).$ The values of the fields 
on the two sides of the shock are related by the Rankine-Hugoniot conditions:
\be  \Delta (\rho v)=0, \quad \Delta(\frac{1}{2}v^2+h_m)=0, \quad \Delta (p+\rho v^2)=0, \lb{eqA6} \ee
with a mass flux $j_*=\rho_0 v_0=\rho_1 v_1>0$. 
See \cite{landau2013fluid}, \S 84. The strength of the shock is characterized by the 
compression ratio $R=\rho_1/\rho_0=v_0/v_1>1$ which, for an ideal gas, is given by 
\be  R = \frac{\gamma+1}{(\gamma-1)+2/M_0^2} \lb{eqA7} \ee
in terms of the pre-shock Mach number $M_0=v_0/c_s>1.$ E.g. see 
Landau \& Lifschitz (1987), \S 89. Note that, because of the ideal gas relation $p/\rho=(\gamma-1)h_m/\gamma,$
the second two Rankine-Hugoniot conditions determine the pre- and post-shock pressures by the formulas
\be  p_i =\frac{j_*}{2\gamma}[(1+\gamma)v_{1-i} +(1-\gamma)v_i ], \quad i=0,1. \lb{eqA8} \ee

As was first noted by Becker \cite{becker1922stosswelle}, the stationary 1D Navier-Stokes equations of an ideal gas 
admit an exact integral for $\eta=\kappa/c_P$ or, assuming $\zeta_{3D}=0,$ for the 3D Prandtl number 
$Pr=c_P \eta_{3D}/\kappa=3/4.$ This integral takes the form of a (non-ideal) Bernoulli equation which 
relates velocity and enthalpy per mass:
\be \frac{1}{2}v^2+h_m=\frac{1}{2}v^2_0+h_{m0} = \frac{\gamma+1}{2(\gamma-1)}v_0v_1, \quad Pr=3/4.  \lb{eqA9} \ee
By means of this relation and the formula 
\be \rho=j_*/v \lb{eqA10} \ee
for the mass density, all thermodynamic variables can be related to the velocity. 
For example, using (\ref{eqA9}) and the ideal gas relation $p/\rho=(\gamma-1)h_m/\gamma$ gives for the pressure field
\be  p=\frac{j_*}{2\gamma}\left[(\gamma+1)\frac{v_0v_1}{v}+(1-\gamma)v\right], \quad Pr=3/4. \lb{eqA11} \ee
Using $h_m=c_PT$ gives for the temperature field 
\be  T=\frac{1}{2c_P}\left[\frac{\gamma+1}{\gamma-1}v_0v_1-v^2\right], \quad Pr=3/4, \lb{eqA12} \ee
and likewise for other thermodynamic quantities. For a very clear discussion, see \cite{johnson2013analytical}. 

To obtain the velocity itself in the approach of \cite{becker1922stosswelle} requires the evaluation of an integral 
involving the specific choice of dynamic viscosity $\eta(\rho,T)$ as a function of $\rho$ and $T.$ 
This generally yields the velocity field in the implicit form $x(v).$ As pointed out by Johnson (2014), 
some choices of $\eta(\rho,T)$ permit one to invert the relation $x(v)$ to an explicit form $v(x).$ 
It turns out, however, that to evaluate the infinite Reynolds-number/P\'eclet-number limits, we need 
only the Bernoulli relation (\ref{eqA9}) of \cite{becker1922stosswelle} and its alternative forms (\ref{eqA11}),(\ref{eqA12}). 
We furthermore need one additional constraint which follows from the constancy of momentum flux:
\be   j_* v + p - \eta \partial_x v \equiv \tau_*. \lb{eqA13} \ee
The constant value $\tau_*$ can be evaluated far from the shock where the gradient vanishes, giving
\be  \tau_* = j_* v_i + p_i, \quad i=0,1 \lb{eqA14} \ee 
or, using (\ref{eqA8}) from the Rankine-Hugoniot conditions, 
\be  \tau_* = j_* v_{av} + p_{av}= \frac{1+\gamma}{\gamma} j_* v_{av}. \lb{eqA15} \ee
The equations (\ref{eqA13}),(\ref{eqA15}) and the Bernoulli relation (\ref{eqA11}) for $p$ allow us to determine $\eta\partial_x v$ in terms 
of $v$ itself, yielding identical results for any choice of viscosity $\eta(\rho,T).$ As a consequence, all of our 
ideal limit results are independent of the details of the molecular transport coefficients, apart from the requirement 
that $Pr=3/4.$ Many inertial-range limit results hold with complete generality for all dissipative 
planar shocks in an ideal gas, and do not even depend upon Prandtl number $Pr.$ Some inertial-range 
quantities do depend upon $Pr$, which we can explicitly verify 
for the cases $Pr=\infty$ ($\kappa=0$) and $Pr=0$ ($\eta=0$). As was noted by \cite{johnson2014closed},
there are Bernoulli-type relations also for those cases, which yield expressions for the pressure of the form
\be  p=j_*\left[ -\frac{1-\gamma}{2}v + \frac{1-\gamma^2}{\gamma} v_{av} + \frac{1+\gamma}{2}\frac{v_0v_1}{v}\right],
\quad Pr=\infty, \lb{eqA16} \ee  
and
\be  p = j_*\left[ \frac{1+\gamma}{\gamma}v_{av}-v\right], \quad Pr=0. \lb{eqA17} \ee
Employing these expressions for $p$ and eqs.(\ref{eqA13}),(\ref{eqA15}) we can also obtain formulas for $\eta\partial_xv$ with
$Pr=0,\infty$ which allow us to extend all of our results for $Pr=3/4$ to those cases. Because the mathematical methods 
are essentially the same for all three cases, we shall below discuss explicitly only $Pr=3/4$ and then just briefly mention 
some corresponding results for $Pr=0,\infty.$ 



These solutions of Becker \cite{becker1922stosswelle} and Johnson \cite{johnson2013analytical,johnson2014closed}
are a nice example 
for our general mathematical framework, since they converge in $L^p$ norms for any $p\in [1,\infty)$ to a 
weak shock solution of 1D compressible Euler as $\nu\rightarrow 0.$  
We derive here all of the source terms which appear in the kinetic energy and the entropy balance for the 
shock solutions in the distributional limit as $\eta,\ \kappa\rightarrow 0$ for the fine-grained balances and as 
$\ell\rightarrow 0$ for the coarse-grained balances.  A fact that we shall use frequently 
for ideal step-function fields below is 
\be  \bar{f}(x) = f_0 + (\Delta f)\bar{\theta}(x), \quad \bar{g}(x) = g_0 + (\Delta g)\bar{\theta}(x) \lb{eqA18} \ee
and thus
\be  \bar{g} = g_0 + \frac{\Delta g}{\Delta f}(\bar{f}-f_0), \quad 
\partial_x \bar{g} =   \frac{\Delta g}{\Delta f}\partial_x \bar{f}. \lb{eqA19} \ee
Furthermore, 
\be  \partial_x\bar{f}(x) = (\Delta f)\bar{\delta}(x). \lb{eqA20} \ee
Similar results can be obtained from
\be  \bar{f}(x) = f_{av} + \frac{1}{2}(\Delta f)\overline{{\rm sign}}(x), \quad 
\bar{g}(x) = g_{av} + \frac{1}{2}(\Delta g)\overline{{\rm sign}}(x) \lb{eqA21} \ee
These relations are very helpful to derive inertial-range expressions for the shock solution. 

\subsection{Kinetic Energy Balance}\label{sec:shockkinetic}

\subsubsection{Viscous Dissipation}\label{sec:shockviscous}

Using in (\ref{eqA13}) expression (\ref{eqA11}) for $p$ with $Pr=3/4$ 
\be \eta (\partial_x v)=\frac{\gamma+1}{2\gamma}j_*\left[v-2v_{av}+\frac{v_0v_1}{v}\right].\lb{eqA22} \ee
Hence, 
\begin{eqnarray}
&& \eta (\partial_x v)^2=\frac{\gamma+1}{2\gamma}j_*\Big[ \partial_x\left(\frac{1}{2}v^2\right)\cr
&& \hspace{70pt}  -2v_{av}\partial_x v+v_0v_1\partial_x(\ln v)\Big].
\lb{eqA23} \end{eqnarray} 
Since 
\begin{eqnarray}
&& \Dlim_{\eta,\ \kappa\rightarrow 0}\left[\partial_x\left(\frac{1}{2}v^2\right), \ \partial_x v, \ \partial_x(\ln v)\right] \cr
&& \hspace{50pt} = \left[\frac{1}{2}(v^2_1-v_0^2), \ \Delta v, \ \ln(v_1/v_0)\right]\delta(x)  \cr
&& \lb{eqA24} \end{eqnarray}
one gets easily that 
\begin{eqnarray}
Q_{visc}&\equiv & \Dlim_{\eta,\ \kappa\rightarrow 0}\  \eta(\partial_x v)^2 \cr
   &=&\frac{\gamma+1}{2\gamma}j_*\Big[ v_1v_0\ln\left(\frac{v_1}{v_0}\right) -\frac{1}{2}(v_1^2-v_0^2)\Big]\delta(x) \cr
&& \lb{eqA25} \end{eqnarray} 
Note that $Q_{visc}\geq 0$ because 
\be  f(\theta)= \theta\ln \theta -\frac{1}{2}\theta^2+\frac{1}{2}>0, \qquad 0\leq \theta<1 \lb{eqA26} \ee
and $f(1)=0.$ This result is Prandtl-number dependent. In fact, $Q_{visc}$ for $Pr=\infty$ is larger 
by factor of $\gamma$ and for $Pr=0,$ obviously, $Q_{visc}=0.$ 

\subsubsection{Pressure-Dilatation Defect}\label{sec:shockpTheta}

From (\ref{eqA11}) for $p$ with $Pr=3/4$ we have  
\begin{eqnarray}
p(\partial_xv)&=& \frac{j_*}{2\gamma}\left[(\gamma+1)\frac{v_0v_1}{v}-(\gamma-1)v\right]\partial_xv \cr
&=& j_*\frac{\gamma+1}{2\gamma}v_0v_1\partial_x(\ln v) -  j_*\frac{\gamma-1}{2\gamma}\partial_x\left(\frac{1}{2}v^2\right) \cr
&& \lb{eqA27} \end{eqnarray}
Thus,
\begin{eqnarray}
&& p*\Theta \equiv  \Dlim_{\eta,\ \kappa\rightarrow 0}\  p(\partial_x v)  \cr
   &&=\frac{j_*}{2\gamma}\Big[ (\gamma+1)v_1v_0\ln\left(\frac{v_1}{v_0}\right) -\frac{1}{2}(\gamma-1)(v_1^2-v_0^2)\Big]\delta(x) \cr
&& \lb{eqA28} \end{eqnarray} 
This result is also Prandtl-number dependent (see below) and the above expression holds only for $Pr=3/4$. 

Next we calculate $p\circ \Theta.$ Since $v$ and $p$ in the ideal limit are both step functions, $\partial_x\bar{v} = (\Delta v/\Delta p)\partial_x\bar{p}$
so that 
\be  \bar{p}\partial_x\bar{v} =\frac{\Delta v}{\Delta p}\partial_x\left(\frac{1}{2}\bar{p}^2\right)\lb{eqA29} \ee
and thus 
\begin{eqnarray} 
p\circ\Theta &\equiv&  \Dlim_{\ell\rightarrow 0}\  \bar{p}(\partial_x \bar{v}) \cr
   &=&(\Delta v)p_{av}\delta(x) = \frac{j_*}{2\gamma}\left(v_1^2-v_0^2\right)\delta(x)
\lb{eqA30} \end{eqnarray} 
using $p_{av}=j_*v_{av}/\gamma$ from eq.(\ref{eqA8}). Note that this result is independent of the particular choice 
of filter kernel $G,$ as required. It is also completely independent of the molecular dissipation, as it is determined 
solely from the limiting Euler solution fields.  One finds by subtracting that 
\begin{eqnarray}
&& \tau(p,\Theta)\equiv  p*\Theta-p\circ \Theta  \cr
&&\hspace{2pt} =\frac{\gamma+1}{2\gamma}j_*\Big[ v_1v_0\ln\left(\frac{v_1}{v_0}\right) -\frac{1}{2}(v_1^2-v_0^2)\Big]\delta(x)
\lb{eqA31} \end{eqnarray} 
Clearly, $Q_{visc}=\tau(p,\Theta)$ for $Pr=3/4.$ This same identity in fact holds for all values of Prandtl number,
allowing us to infer the $Pr$-dependence of $p*\Theta$ from that of $Q_{visc}.$ The underlying reason for this 
identity, which is valid for all planar shocks in an ideal gas, is explained in the next subsection on kinetic energy flux.  

\subsubsection{Kinetic Energy Flux}\label{sec:shockenergyflux}

{\it Baropycnal work:} Using $\bar{\tau}(\rho,v)=\overline{\rho v}-\bar{\rho}\bar{v}=j_*-\bar{\rho}\bar{v},$
\be  \frac{1}{\bar{\rho}}\bar{\tau}(\rho,v)= \frac{j_*}{\bar{\rho}}-\bar{v}. \lb{eqA32} \ee
Using $\partial_x\bar{p} = (\Delta p/\Delta \rho)\partial_x\bar{\rho}= (\Delta p/\Delta v)\partial_x\bar{v},$
\be  \frac{1}{\bar{\rho}}\bar{\tau}(\rho,v) \partial_x\bar{p} = j_*\frac{\Delta p}{\Delta\rho}\partial_x(\ln\bar{\rho})
-\frac{\Delta p}{\Delta v} \partial_x\left(\frac{1}{2}\bar{v}^2\right). \lb{eqA33} \ee
Thus,
\begin{eqnarray}
Q_{baro} &\equiv & \Dlim_{\ell\rightarrow 0}  \frac{\partial_x\bar{p}}{\bar{\rho}}\bar{\tau}(\rho,v)   \cr
   &=&\Delta p\Big[ \frac{j_*}{\Delta \rho}\ln\left(\frac{\rho_1}{\rho_0}\right) -v_{av}\Big]\delta(x)\cr
   &=& -j_* \Big[ v_1v_0\ln\left(\frac{v_1}{v_0}\right) -\frac{1}{2}(v_1^2-v_0^2)\Big]\delta(x)
\lb{eqA34} \end{eqnarray} 
where the final line was obtained using $\Delta p=-j_*\Delta v,$ which follows either from (\ref{eqA8}) or directly
from the Rankine-Hugoniot conditions (\ref{eqA6}). Note that $Q_{baro}\leq 0.$ 

We see again that 
the limiting inertial-range result $Q_{baro}$ is independent of the filter kernel $G.$ This is true for all of the limits 
as $\ell\rightarrow 0$ of inertial-range expressions for the shock solutions that we obtain in this Appendix. 
Thus, we shall make no further note of this fact for the other limits derived below. Note that $Q_{baro}$
is also completely independent of the molecular transport coefficients, as are all other quantities 
that are determined solely by the limiting Euler solution fields. 

\vspace{5pt} 

{\it Deformation Work:} Using $\tilde{v}=\overline{\rho v}/\bar{\rho}=j_*/\bar{\rho}$
\be  \bar{\rho}\partial_x\tilde{v} = -\frac{j_*}{\bar{\rho}}\partial_x\bar{\rho}. \lb{eqA35} \ee
Likewise from its definition and $\rho v=j_*$ one gets
\begin{eqnarray}
&& \tilde{\tau}(v,v) =\frac{j_*\bar{v}}{\bar{\rho}}-\frac{j_*^2}{\bar{\rho}^2}\cr
&& =j_*\left[\frac{\Delta v}{\Delta \rho}
+\left(v_{av}-\frac{\Delta v}{\Delta \rho}\rho_{av}\right) \frac{1}{\bar{\rho}}-\frac{j_*}{\bar{\rho}^2}\right]
\lb{eqA36} \end{eqnarray} 
after substituting $\bar{v}=v_{av}+\frac{\Delta v}{\Delta \rho}(\bar{\rho}-\rho_{av})$ from (\ref{eqA21}). Thus,
\begin{eqnarray}
&& Q_{defor} \equiv  - \Dlim_{\ell\rightarrow 0}\ \bar{\rho}(\partial_x\tilde{v})\tilde{\tau}(v,v)   \cr
   && =j_*^2 \left[\frac{\Delta v}{\Delta \rho}\Delta(\ln \rho)
   -\left(v_{av}-\frac{\Delta v}{\Delta \rho}\rho_{av}\right)\Delta(\frac{1}{\rho}) \right. \cr
   && \hspace{120pt} \left.
   + j_*\Delta(\frac{1}{2\rho^2})\right]\delta(x) \cr
      &=& j_* \Big[ v_1v_0\ln\left(\frac{v_1}{v_0}\right) -\frac{1}{2}(v_1^2-v_0^2)\Big]\delta(x)\geq 0
\lb{eqA37} \end{eqnarray} 
upon simplification. Thus, $Q_{flux}=Q_{baro}+Q_{defor}=0.$ \\ Since $Q_{visc}=\tau(p,\Theta)+Q_{flux}$ 
in general, this explains why $Q_{visc}=\tau(p,\Theta)$ holds independent of the molecular transport coefficients
for any planar, ideal-gas shock. Note that the last identity can also be restated as 
$-p*\Theta +Q_{visc}=-p\circ \Theta,$ which corroborates for these solutions the general argument in the text 
that the sum of  $-p*\Theta$ and $Q_{visc}$ should be completely independent of the molecular dissipation, 
even though the two terms separately are Prandtl-number dependent.  

\vspace{5pt} In physical terms, there is a loss of kinetic energy $-p\ \circ \Theta$ at the shock, and an 
equal gain $-p*\Theta+Q_{visc}$ of internal energy. There is no external forcing to balance 
the kinetic energy loss and no cooling to balance the internal energy gain. While these shock solutions are 
stationary, they are not however homogeneous or isotropic. Thus, the loss/gain is balanced by space-transport of kinetic/internal 
energy into/away from the shock. For example, the space flux of kinetic energy is 
\be  J_{kin}=\left(\frac{1}{2}\rho v^2 + p\right)v=  j_*\left(\frac{1}{2}v^2 + \frac{p}{\rho}\right). \lb{eqA38} \ee
One readily finds from eq.(\ref{eqA8}) for $p_i,$ $i=0,1$ that
\be  \Delta J_{kin} = \frac{j_*}{2\gamma}(v_1^2-v_0^2)<0 \lb{eqA39} \ee
so that more kinetic energy enters the shock than leaves it, and the difference is exactly the correct amount 
to offset the loss due to pressure-work. Similarly, more internal energy is transported away from the shock 
than enters it, balancing the gain from pressure-work and heating. This follows directly from the conservation 
of total energy, or else by using $J_{int}= uv$ for space-flux of internal energy and evaluating 
$\Delta J_{int}=-\Delta J_{kin}.$

\subsection{Entropy Balance}\label{sec:shockentropy} 

\subsubsection{Dissipation Range}\label{sec:shockdissipation} 

{\it Viscous Heating} $\beta*Q_{visc}$:  Using (\ref{eqA22}) for $\eta\partial_x v$ write
\begin{eqnarray}
&& \frac{\eta (\partial_x v)^2}{T}=\frac{\gamma+1}{2\gamma}j_*\left[v-2v_{av}+\frac{v_0v_1}{v}\right]\frac{\partial_x v}{T}\cr
&& = \frac{\gamma+1}{2\gamma}j_*\left(1+\frac{v_0v_1}{v^2}\right)\frac{v\partial_x v}{T}
        -\frac{\gamma+1}{\gamma}j_* v_{av}\frac{\partial_x v}{T} \cr
&& \lb{eqA40} \end{eqnarray}
In the first term replace $v^2$ with $T$ using both (\ref{eqA12}) and its derivative $v\partial_xv=-c_P\partial_x T,$ while in 
the second term replace $T$ with $v^2$ using (\ref{eqA12}). Elementary anti-derivatives give for the first term 
\be  -j_* \gamma c_V \partial_x(\ln T)+\frac{1}{2}j_*(\gamma-1) c_V \partial_x \ln|a^2-2c_P T|\lb{eqA41} \ee
with $a^2=\frac{\gamma+1}{\gamma-1}v_0v_1$
and for the second term
\be  -2j_* c_V (\gamma+1) \frac{v_{av}}{a}\partial_x {\rm arctanh}(v/a) \lb{eqA42} \ee
Noting that $a^2-2c_P T= v^2=j_*^2/\rho^2$ gives finally that 
\begin{eqnarray}
&& \beta*Q_{visc} = \Dlim_{\eta,\ \kappa\rightarrow 0} \frac{\eta (\partial_x v)^2}{T} \cr
&& \hspace{20pt} = -j_* c_V\Big[\gamma \ln\left(\frac{T_1}{T_0}\right)+(\gamma-1) \ln\left(\frac{\rho_1}{\rho_0}\right) \cr
&& \hspace{30pt} + 2(\gamma+1)\frac{v_{av}}{a}\Delta({\rm arctanh}(v/a))  \Big]\delta(x).
\lb{eqA43} \end{eqnarray}
This expression holds only for $Pr=3/4$ and the quantity is generally Prandtl-number dependent. For example,
$\beta*Q_{visc}=0$ for $Pr=0.$ 

\vspace{5pt} 

{\it Thermal Conduction} $\Sigma_{therm}$: Note since $Pr=3/4$ and $v\partial_x v=-c_P\partial_x T$ that 
\be   \kappa\partial_x T = -\eta v\partial_x v, \lb{eqA44} \ee
and thus also using (\ref{eqA22}) for $\eta\partial_x v$ that 
\begin{eqnarray}
\kappa\partial_x T
&=& - j_*\frac{\gamma+1}{2\gamma}(v^2-2v_{av}v+v_0v_1) \cr
&=& -j_*(\gamma+1)\left(\frac{v_0v_1}{\gamma-1}-c_V T\right) + j_* \frac{\gamma+1}{\gamma}v_{av}v. \cr
&& \lb{eqA45} \end{eqnarray}
Hence,
\begin{eqnarray}
 \frac{\kappa (\partial_x T)^2}{T^2} &=&  -j_*(\gamma+1)\left(\frac{v_0v_1}{\gamma-1}-c_V T\right) \frac{\partial_x T}{T^2} \cr 
&& \hspace{20pt} - 4j_* c_V(\gamma+1) v_{av} \frac{v^2\partial_x v}{(a^2-v^2)^2}
\lb{eqA46} \end{eqnarray}
Elementary anti-derivatives and some lengthy algebraic simplifications give
\begin{eqnarray}
&&  \frac{\kappa (\partial_x T)^2}{T^2} =  j_* \frac{\gamma+1}{\gamma-1}v_0v_1\partial_x\left(\frac{1}{T}\right) + j_* c_V(\gamma+1)\partial_x (\ln T)\cr 
&&- j_* \frac{\gamma+1}{\gamma} v_{av} \partial_x\left(\frac{v}{T}\right)+ 2j_*c_V(\gamma+1) \frac{v_{av}}{a}\partial_x {\rm arctanh}(v/a) \cr
&& \lb{eqA47} \end{eqnarray}
Taking the limit $\eta,\ \kappa\rightarrow 0,$ one finds using the Bernoulli relation that the contributions of the first and third terms
in the above expression cancel, giving the final result
\begin{eqnarray}
&& \Sigma_{therm} = \Dlim_{\eta,\ \kappa\rightarrow 0}  \frac{\kappa (\partial_x T)^2}{T^2}\cr 
&& = j_* c_V (\gamma+1) \left[\ln\left(\frac{T_1}{T_0}\right) + 2\frac{v_{av}}{a}\Delta({\rm arctanh}(v/a)) \right] \delta(x). \cr
&& \lb{eqA48} \end{eqnarray} 
Once again, this quantity is Prandtl-number dependent and the above expression holds only for $Pr=3/4.$ 
Obviously  $\Sigma_{therm}=0$ for $Pr=\infty.$ 

\vspace{5pt} 

{\it Total Entropy Production} $\Sigma_{diss}$: The inverse hyperbolic tangent terms cancel on addition, giving 
\begin{eqnarray}
\Sigma_{diss} &=&  \Sigma_{therm} + \beta* Q_{visc} \cr 
&=& j_* c_V  \left[\ln\left(\frac{T_1}{T_0}\right) -(\gamma-1) \ln\left(\frac{\rho_1}{\rho_0}\right)\right] \delta(x) \cr
&=& j_* \Delta s_m \delta(x),  
\lb{eqA49} \end{eqnarray}
using $s_m=c_V\ln(T/C\rho^{\gamma-1}).$ We see that $\Sigma_{diss}>0,$ since $\Delta s_m>0$
is the standard entropy condition for an Euler shock.  The result (\ref{eqA49}) could have been anticipated on the basis of 
simple entropy balance, since $J_{ent}=s v = s_m j_*$ is the space-flux of entropy and $\Delta J_{ent}=(\Delta s_m)j_*$
is the net entropy transported away from the shock. Thus, the entropy production at the shock is balanced 
by transport of entropy to infinity. The result (\ref{eqA49}) for $\Sigma_{diss}$ is, for this reason, completely independent
of the molecular dissipation. Note that $\Sigma_{therm}=\Sigma_{diss}$ for $Pr=0$ and that 
$\beta*Q_{visc}=\Sigma_{diss}$ for $Pr=\infty.$ These results for $Pr=0,\infty$ can be obtained as well using the Bernoulli-type
relations in \cite{johnson2013analytical,johnson2014closed} and calculating in the same manner as for $Pr=3/4$ above. 

\subsubsection{Inertial-Range}\label{sec:shockinertial} 

{\it Inertial-Range Viscous Heating} $\beta\circ Q_{visc}:$ From (\ref{eqA25}), $Q_{visc}=q_*\delta (x),$ so that 
\be  \bar{Q}_{visc} = q_*\bar{\delta}(x). \lb{eqA50} \ee
On the other hand, for an ideal gas by definition of $\ubeta$
\be  \ubeta= c_V \frac{\bar{\rho}}{\bar{u}}. \lb{eqA51} \ee
Because $u,\rho$ are step-functions in the ideal limit,
\be \bar{\rho}=\left(\rho_0-\frac{\Delta\rho}{\Delta u}u_0\right)+\frac{\Delta\rho}{\Delta u}\bar{u},
\qquad \bar{\delta}= \frac{\partial_x\bar{u}}{\Delta u}.\lb{eq52} \ee
Thus,
\begin{eqnarray}
 \ubeta\bar{Q}_{visc} &=& c_V q_*\left[
  \left(\rho_0-\frac{\Delta\rho}{\Delta u}u_0\right)\frac{\partial_x\bar{u}}{\bar{u}\Delta u}.
 +\frac{\Delta\rho}{\Delta u}\bar{\delta}(x)\right] \cr
&=& c_V q_*\left[\left(\frac{\rho_0\Delta u-u_0\Delta\rho}{(\Delta u)^2}\right)
\partial_x(\ln\bar{u})+\frac{\Delta\rho}{\Delta u}\bar{\delta}(x)\right] \cr
&& \lb{eqA53} \end{eqnarray} 
Hence 
\be 
\beta\circ Q_{visc} = \Dlim_{\ell\rightarrow 0}\  \ubeta\bar{Q}_{visc} =\beta_* q_*\delta(x) \lb{eq54} \ee
with 
\be  \beta_*\equiv c_V \left[\left(\frac{\rho_0 u_1-u_0\rho_1}{(\Delta u)^2}\right)
\ln\left(\frac{u_1}{u_0}\right)+\frac{\Delta\rho}{\Delta u}\right] \lb{eqA55} \ee
This result is obviously independent of the filter-kernel $G$ (as are all such limits of 
coarse-grained quantities), but the quantity $q_*$ gives a Prandtl-number dependence. 

\vspace{5pt} 

{\it Pressure-Dilatation Defect} $\beta\circ\tau(p,\Theta):$
Because of our earlier result $p*\Theta=q_{PV}\delta(x),$ the same argument as above shows that
\be  \Dlim_{\ell\rightarrow 0}\ \ubeta\overline{p*\Theta} = \beta_* q_{PV} \delta(x). \lb{eqA56} \ee
Next note using $\bar{p}=(\gamma-1)\bar{u}$ and (\ref{eqA51}) for $\ubeta$ that
\be  \ubeta\ \bar{p}\ \partial_x\bar{v}= c_V(\gamma-1) \bar{\rho}\partial_x\bar{v}
= c_V(\gamma-1)\frac{\Delta v}{\Delta\rho}\partial_x\left(\frac{1}{2}\bar{\rho}^2\right), \lb{eqA57} \ee
where the last equality follows from $\partial_x\bar{v}=(\Delta v/\Delta \rho)\partial_x\bar{\rho}.$
Thus, 
\begin{eqnarray}
\Dlim_{\ell\rightarrow 0}\ \ubeta \ \bar{p}\ \partial_x\bar{v} &=& c_V(\gamma-1) (\Delta v)\rho_{av} \delta(x) \cr
&=& c_Vj_*(\gamma-1) \frac{(\Delta v)v_{av}}{v_1v_0} \delta(x) \cr
&& \lb{eqA58} \end{eqnarray}
after using $\rho=j_*/v.$ Finally,
\be  \beta\circ\tau(p,\Theta) = \left[\beta_*q_{PV}-c_Vj_*(\gamma-1) \frac{(\Delta v)v_{av}}{v_1v_0} \right]\delta(x).\lb{eqA59} \ee
This quantity is of course Prandtl-number dependent through the coefficient $q_{PV}$. 

\vspace{5pt} 

{\it Combined Contribution} $\beta\circ Q-\beta\circ\tau(p,\Theta):$
Using the expression (\ref{eqA25}) for $q_*$ and (\ref{eqA28}) for $q_{PV},$ one can see that the log-term cancels in the difference and
\be  q_*-q_{PV}= -\frac{j_*}{\gamma} \Delta \left(\frac{1}{2}v^2\right) = j_* c_V (\Delta T). \lb{eqA60} \ee
In accord with our earlier remarks, this is the same as the coefficient of $-p\circ \Theta$ and is completely 
independent of choice of molecular transport coefficients.  Thus, 
\be  \beta\circ Q-\beta\circ\tau(p,\Theta) 
= c_V j_* \left[\beta_* (\Delta T) +(\gamma-1)\frac{v_{av}\Delta v}{v_0v_1}\right]\delta(x) \lb{eqA61} \ee
and is also independent of molecular dissipation. 

\vspace{5pt} 

{\it Negentropy Flux} $\Sigma_{flux}:$
We first consider the contribution from $(\partial_x\ubeta)\overline{uv}.$ From $\ubeta= c_V \bar{\rho}/\bar{u}$ one gets
\be  \partial_x\ubeta = c_V\left(\frac{1}{\bar{u}}\partial_x\bar{\rho}-\frac{\bar{\rho}}{\bar{u}^2}\partial_x\bar{u}\right)\lb{eqA62} \ee
whereas 
\be  \overline{uv} = c_V \overline{\rho Tv} = c_V j_* \overline{T}. \lb{eqA63} \ee
Writing $\bar{\rho}$ in terms of $\bar{u}$ using (\ref{eqA19}) and the similar relation for $\bar{T}$ in terms of $\bar{u},$
one finds after some simplifications 
\begin{eqnarray}
&& (\partial_x\ubeta)\overline{uv} = -c_V^2 j_*\left(\rho_0-\frac{\Delta \rho}{\Delta u} u_0\right) \cr 
&& \hspace{50pt} \times \left[\left(T_0-\frac{\Delta T}{\Delta u} u_0\right)\frac{\partial_x\bar{u}}{\bar{u}^2} 
+ \frac{\Delta T}{\Delta u} \frac{\partial_x\bar{u}}{\bar{u}}\right] 
\lb{eqA64} \end{eqnarray} 
Thus,
\begin{eqnarray}
&& \Dlim_{\ell\rightarrow 0} \ (\partial_x\ubeta)\overline{uv} = -c_V^2 j_*\left(\rho_0-\frac{\Delta \rho}{\Delta u} u_0\right) \cr 
&& \hspace{50pt} \times \left[\left(T_0-\frac{\Delta T}{\Delta u} u_0\right)\frac{\Delta u}{u_0u_1}+ 
\frac{\Delta T}{\Delta u} \ln\left(\frac{u_1}{u_0}\right)\right] \delta(x). \cr
&& \lb{eqA65} \end{eqnarray} 
Using the relation 
\be  (T_0u_1-T_1u_0)(\rho_0u_1-\rho_1u_0)=(\Delta T)(\Delta \rho)u_0 u_1\lb{eqA66} \ee
that follows from $u=c_V\rho T,$ and the definition of $\beta_*$ from (\ref{eqA55}), this reduces to 
\be \Dlim_{\ell\rightarrow 0}\ \partial_x\ubeta\cdot\overline{uv} = -\beta_* c_V (\Delta T) j_*\delta(x) \lb{eqA67} \ee

Next note that $\partial_x \ulambda_m\cdot\overline{\rho v} = j_* \partial_x \ulambda_m$. Hence, 
\be \Dlim_{\ell\rightarrow 0}\ \partial_x \ulambda_m\cdot\overline{\rho v} = (\Delta \lambda_m)j_*\delta(x). \lb{eqA68} \ee
For any equation of state, the Gibbs fundamental relation may be written as $s_m=h_m/T-\lambda_m.$
For an ideal gas $h_m = c_P T, $ so that $s_m=c_P-\lambda_m$ and $\Delta s_m =-(\Delta \lambda_m).$
Thus, 
\be \Dlim_{\ell\rightarrow 0}\ \partial_x \ulambda_m\cdot\overline{\rho v} = - (\Delta s_m)j_*\delta(x)\lb{eqA69} \ee

Finally, consider the contributions from $\partial_x\ubeta(\bar{u}\cdot \bar{v})$, $\partial_x \ulambda_m(\bar{\rho}\cdot \bar{v}).$
Using the equation (\ref{eqA62}) for $\partial_x\ubeta$ one has
\be  \partial_x\ubeta(\bar{u}\cdot \bar{v}) = c_V \bar{v} (\partial_x\bar{\rho}) - c_V \frac{\bar{\rho}\bar{v}}{\bar{u}}(\partial_x\bar{u})\lb{eqA70} \ee
Next use $\partial_x\ulambda_m=-\partial_x\us_m$ and $s_m=c_V\ln(u/C\rho^\gamma)$ to obtain
\be  \partial_x\ulambda_m (\bar{\rho}\cdot \bar{v})= c_V \gamma \bar{v} (\partial_x\bar{\rho}) - c_V \frac{\bar{\rho}\bar{v}}{\bar{u}}(\partial_x\bar{u}).\lb{eqA71} \ee
Subtracting these two expressions gives
\begin{eqnarray}
&& \partial_x\ubeta(\bar{u}\cdot \bar{v}) - \partial_x \ulambda_m(\bar{\rho}\cdot \bar{v}) \cr
&& \qquad =c_V  (\gamma-1) \bar{v} (\partial_x\bar{\rho}) =  (\gamma-1) c_V \frac{\Delta\rho}{\Delta v} \bar{v} (\partial_x\bar{v}). \cr
&& \lb{eqA72} \end{eqnarray} 
From this it follows easily that 
\begin{eqnarray}
&& \Dlim_{\ell\rightarrow 0}\left[ \partial_x\ubeta(\bar{u}\cdot \bar{v}) - \partial_x \ulambda_m(\bar{\rho}\cdot \bar{v})\right] \cr
&& \hspace{90pt} =  c_V(\gamma-1)j_* \frac{v_{av}\Delta v}{v_0v_1}\delta(x). 
\lb{eqA73} \end{eqnarray} 

Putting together all of these results,
\begin{eqnarray}
&& \Sigma_{flux} = \Dlim_{\ell\rightarrow 0}\left[ \partial_x\ubeta \bar{\tau}(u,v) - \partial_x \ulambda_m \bar{\tau}(\rho,v)\right] \cr
&& =  j_* \left[\Delta s_m-c_V \beta_* (\Delta T)-c_V(\gamma-1)\frac{v_{av}\Delta v}{v_0v_1}\right]\delta(x) \cr
&& \lb{eqA74} \end{eqnarray} 
which is valid in general for the limiting Euler solution independent of microscopic dissipation mechanism. \textcolor{black}{Note 
that the first and third terms in the square bracket of the last expression are positive, while the second term is negative.
We do not present details here, but it is possible to show that the sum of all three terms is strictly positive 
as a function of compression factor $R=\rho_1/\rho_0=v_0/v_1$ and maximum compression 
factor $R_\infty=(\gamma+1)/(\gamma-1)$ over the allowed range $1\leq R\leq R_\infty.$} 

\vspace{5pt} 

{\it Total Entropy Production} $\Sigma_{inert}$: 
For the final inertial-range entropy production we get from (\ref{eqA61}) and (\ref{eqA74}) that 
\be  \Sigma_{inert} =\Sigma_{flux}+\beta\circ (Q-\tau(p,\Theta)) =j_* \Delta s_m\delta(x), \lb{eqA75} \ee
which is independent of the molecular dissipation and in exact agreement with 
the net result of the dissipation-range/fine-grained calculation in section \ref{sec:shockdissipation}.  
\textcolor{black}{From relation (\ref{eq88}),  it is also true that 
\be  \Sigma_{flux*} =\Sigma_{flux}+\beta\circ (Q-\tau(p,\Theta)) =j_* \Delta s_m\delta(x). \lb{eqA75} \ee
The intrinsic negentropy flux consistently gives the net entropy production for this problem, since $I_{flux}=0$ 
for an ideal-gas equation of state.}

\bibliographystyle{apsrev4-1}
\bibliography{bibliography.bib}

\end{document}